\documentclass[a4paper,fleqn,usenatbib,useAMS]{mnras}

\usepackage{graphicx}	
\usepackage{amsmath}	
\usepackage{amssymb}	
\usepackage{multicol}        
\usepackage{bm}		
\usepackage{pdflscape}	

\newcommand{\dnu}{\Delta \nu } 
\newcommand{\numax}{\nu_{\rm max} } 
\newcommand{\kepler}{\emph{Kepler} }
\newcommand{\basta}{\texttt{BASTA} }
\newcommand{\change}[1]{#1}

\usepackage[T1]{fontenc}
\usepackage{ae,aecompl}
\usepackage[toc,page]{appendix}

\usepackage{newtxtext,newtxmath}

\title[Surface correction with the LEGACY sample]{Surface correction of main sequence solar-like oscillators with the \emph{Kepler} LEGACY sample}

\author[Compton et al.]{D. L. Compton$^{1,2}$\thanks{E-mail: d.compton@physics.usyd.edu.au}, T. R. Bedding$^{1,2}$, W. H. Ball$^{3,2}$, D. Stello$^{4}$, D. Huber$^{5,1,2,6}$, \newauthor T. R. White$^{2}$, H.~Kjeldsen$^{2}$ \\
$^{1}$ Sydney Institute for Astronomy, School of Physics, University of Sydney, NSW 2006, Australia \\
$^{2}$ Stellar Astrophysics Centre, Department of Physics and Astronomy, Aarhus University, Ny Munkegade 120, DK-8000 Aarhus C, Denmark \\
$^{3}$ School of Physics and Astronomy, University of Birmingham, Edgbaston, Birmingham, B15 2TT, UK \\
$^{4}$ School of Physics, University of New South Wales, NSW 2052, Australia \\
$^{5}$ Institute for Astronomy, University of Hawai'i, 2680 Woodlawn Drive, Honolulu, HI 96822, USA \\
$^{6}$ SETI Institute, 189 Bernardo Avenue, Mountain View, CA 94043, USA}

\date{Accepted 2018 June 18. Received 2018 June 16; in original form 2018 May 3}

\pubyear{2018}

\begin{document}
\label{firstpage}
\pagerange{\pageref{firstpage}--\pageref{lastpage}}
\maketitle
%
\begin{abstract}
Poor modelling of the surface regions of solar-like stars causes a systematic discrepancy between the observed and model pulsation frequencies. We aim to characterise this frequency discrepancy for main sequence solar-like oscillators for a wide range of initial masses and metallicities. We fit stellar models to the observed mode frequencies of the 67 stars, including the Sun, in the \emph{Kepler} LEGACY sample, using three different empirical surface corrections. The three surface corrections we analyse are a frequency power-law, a cubic frequency term divided by the mode inertia, and a linear combination of an inverse and cubic frequency term divided by the mode inertia. We construct a grid of stellar evolution models using the stellar evolution code MESA and calculate mode frequencies using GYRE. \change{We scale the frequencies of each stellar model by an empirical calculated homology coefficient}, which greatly improves the robustness of our grid. \change{We calculate stellar parameters and surface corrections for each star} using the average of the best-fitting models from each evolutionary track, weighted by the likelihood of each model. The resulting model stellar parameters agree well with an independent reference, the \texttt{BASTA} pipeline. \change{However, we find that the adopted physics of the stellar models has a greater impact on the fitted stellar parameters than the choice of correction method.} We find that scaling the frequencies by the mode inertia improves the fit between the models and observations. The inclusion of the inverse frequency term produces substantially better model fits to lower surface gravity stars.

\end{abstract}




\section{Introduction}
\label{sec:intro}

For the Sun and other solar-like stars, there exists a discrepancy between the observed and modelled frequencies of stellar oscillations. The differences are the consequence of poorly modelled physics at the stellar surface known collectively as the surface effects. A surface correction is routinely applied to the oscillation frequencies of the stellar models to remove this bias, often calculated using an empirical relation between the observed and modelled frequencies for a given stellar model. A number of such relations have been put forward to generalise the correction for all solar-like oscillators. \cite{kjeldsen08} used a frequency power-law to describe the frequency dependence of the correction, which was calibrated based on solar models and data \citep[see][respectively]{cd96,lazrek97}. \cite{ball14} considered two new formulations based on the work by \cite{gough90}: a cubic term, and a linear combination of an inverse and a cubic frequency term. Both methods were scaled by the mode inertia. \cite{schmittbasu15} compared the \cite{kjeldsen08} and \cite{ball14} methods using simulations, rather than observed data, and recommended the latter for any asteroseismic analysis. \cite{sonoi15} proposed another correction function using a modified Lorentzian. \change{Most recently, \cite{nsamba18} used the lower-mass stars in the LEGACY sample when they investigated the systematics that emerge from varying input physics, including the surface correction}. Attempts to characterize the surface correction between observed and simulated mode frequencies of main-sequence stars have so far been limited to the Sun and Sun-like stars, and then extrapolated to the hotter main-sequence stars. These surface correction methods have also been tested on subgiant and red giant solar-like oscillators \citep[e.g.][]{ball17, li18, ball18}.

An alternative approach has been to model the surface of a star using 3D hydrodynamical simulations \citep[e.g.][]{ludwig09, beeck13, trampedach13}. Comparing the difference in pulsation frequencies to the traditional 1D stellar models \citep[e.g.][]{sonoi15,ball16,houdek17} has been used to estimate the required correction, assuming the 3D simulations can more accurately model the surface physics. However, these methods are still incomplete with various components of the surface effect being neglected in the 3D hydrodynamical simulations. Another downside is that the 3D simulations are far more computationally demanding, making them impractical for ensemble stellar analysis. Therefore, there is a desire to find a comprehensive and permanent solution to the erroneous oscillation frequencies calculated from 1D stellar models.

In this project we aimed to assess a number of surface corrections method on an ensemble of main-sequence \kepler stars, named the LEGACY sample \citep{lund17,silva17}. For each star, \cite{silva17} reported the fundamental stellar parameters calculated using seven different pipeline methods. The pipelines adopted a range of surface correction methods, but differing physics and methods of each pipeline made it impossible to properly compare the effect of the surface correction across the different pipeline results for these stars. Therefore, in order to complete such a comparison we also created a pipeline that determined stellar parameters for the LEGACY sample, and the choice of surface correction was adjusted to compare the different functional forms. 

The structure of this paper is as follows: we outline the functional forms of the three proposed surface correction methods in Section~\ref{sec:background}. We also describe how we implemented homology scaling of the oscillations frequencies to increase the robustness of our grid. In Section~\ref{sec:data} we describe how the observed and model data were obtained. Section~\ref{sec:method} is devoted to the method of our pipeline. We explain how we fit the observed and modelled frequencies and calculated the stellar and surface correction parameters for the sample. Results of the analysis and the performance of each surface correction method, as well as their potential shortcomings are discussed in Section~\ref{sec:results}. Finally, conclusions and outlooks are presented in Section~\ref{sec:conc}.





\section{Background}
\label{sec:background}

\subsection{The asymptotic relation}
\label{sec:asym}

The oscillation frequencies of solar-like main-sequence stars \change{approximately} follow an asymptotic relation \citep{tassoul80}
\begin{equation}
\label{equ:asym}
\nu_{n,l} \simeq \dnu \left( n + l/2 + \epsilon \right) - \delta \nu_{0,l},
\end{equation}
where $\dnu$ is the large frequency separation, $n$ is the radial order, $l$ is the angular degree, $\epsilon$ is a phase offset, and $\delta \nu_{0,l}$ is the small separation between modes of different angular degree with respect to the $l=0$ modes. The large frequency separation can be shown to be the inverse of the acoustic travel time through the centre of the star and is approximately proportional to the star's square root mean density $\rho$ \citep{ulrich86,gough87},
\begin{equation}
\label{equ:dnuscale}
\Delta \nu = \left( 2 \int_{0}^{R} \frac{dr}{c} \right)^{-1} \propto \sqrt{\rho},
\end{equation}
where $R$ is the radius of the star and $c$ is the sound speed in the star. The frequency of maximum amplitude,~$\numax$, can also be approximated using fundamental properties of the star,
\begin{equation}
\label{equ:numaxscale}
\nu_\mathrm{max}\propto\frac{g}{\sqrt{T_\mathrm{eff}}}\propto \frac{M}{R^2\sqrt{T_\mathrm{eff}}}
\end{equation}
where $M$ is mass, $T_{\rm eff}$ is the effective temperature, and $g$ is the surface gravity \citep{brown91,kjeldsen95}. The asymptotic relation and the asteroseismic scaling relations can only approximate the properties of solar-like oscillators. Other parameters, such as initial composition and mixing length, must be included to calculate oscillation frequencies of observed stars.
\begin{figure*}
 \includegraphics[width=2\columnwidth]{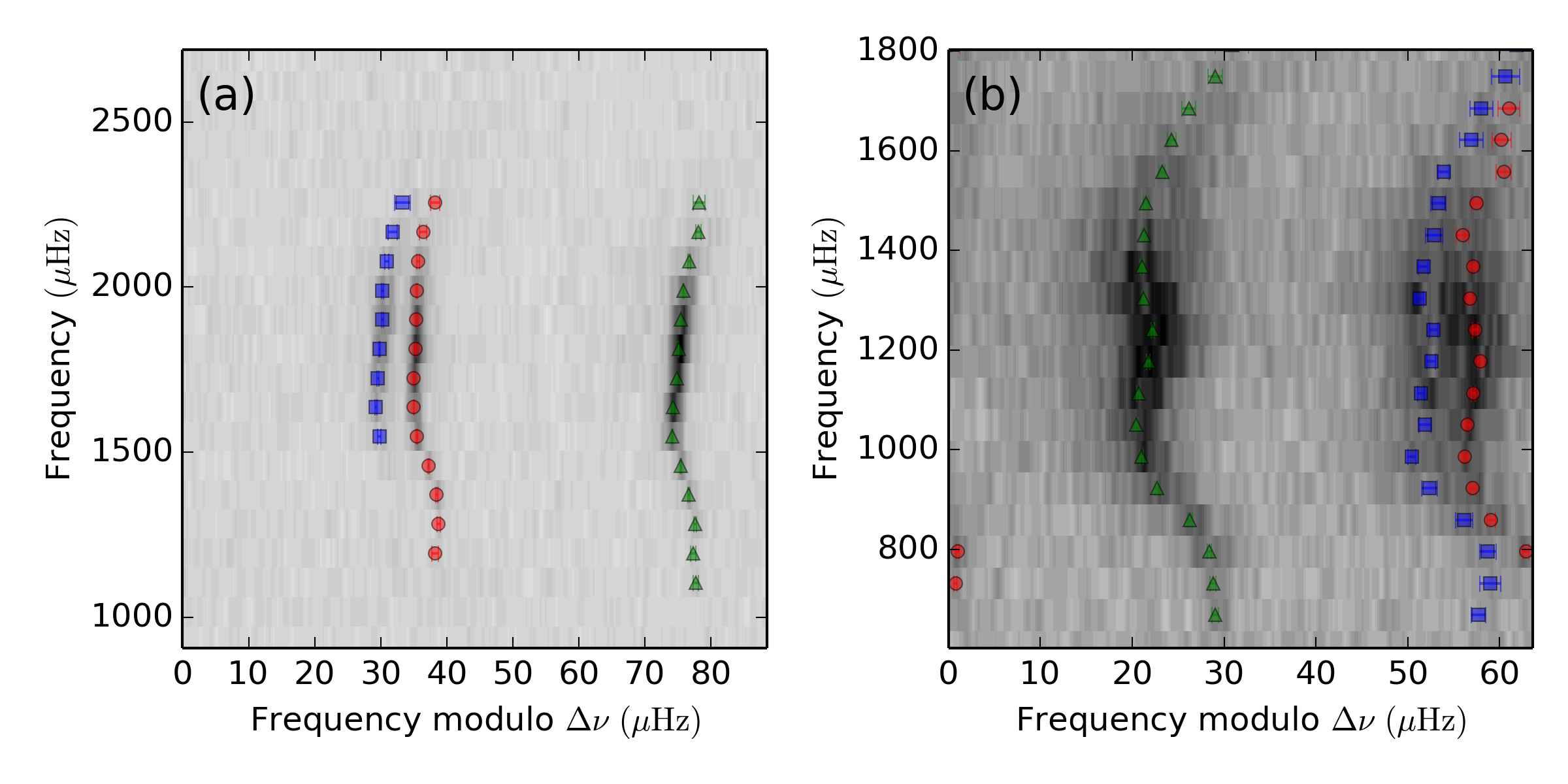}
 \caption{\'{E}chelle diagrams of two \kepler stars in the LEGACY sample. (a) KIC 4914923 and (b) KIC 12317678 show the difference of pulsation power between G and F-type main-sequence stars. The greyscale represents the normalized power of the stellar oscillation spectrum, smoothed with a Gaussian filter with a width of $0.25 \mu{\rm Hz}$. The red circles, green triangles, and blue squares show the extracted frequency peaks \citep[from][]{lund17} for the $l=0$, $l=1$, and $l=2$ angular degrees, respectively. Observational uncertainties are also shown on the corresponding symbols, but are often smaller than the size of the symbol.}
 \label{fig:figure1}
\end{figure*}
Plotting the mode frequencies against the frequencies modulo $\Delta \nu$, called an \'{e}chelle diagram, emphasises departures from the asymptotic relation, that is, the frequency spacing between sequential radial orders is not constant, as implied by Eq.~\ref{equ:asym}. The type and magnitude of these departures from regularity vary depend on the fundamental properties of the star. One example is the higher order curvature that is the result of acoustic glitches due to the oscillations encountering the helium ionisation zone \citep[see][]{verma14b}. Fig.~\ref{fig:figure1}a shows an \'{e}chelle diagram of a \kepler star with similar mass and temperature to the Sun, where the measured frequencies have been plotted on top of the \kepler power spectrum. A more massive main sequence star exhibits more curvature in the \'{e}chelle diagram, shown in Fig.~\ref{fig:figure1}b. This star also shows stronger mode damping, resulting in much broader ridges and lower signal-to-noise.




\subsection{Surface correction}
\label{sec:surfcorr}

To \change{characterize} the discrepancy between the observed frequencies and a best-fitting model, we used an empirically defined surface correction function. In general, the correction increases with oscillation frequency and is largely independent of angular degree for a given star. Consider an observed star with a set of pulsation frequencies, $\nu_{\rm obs} (n, l)$. For this star there exists a model with frequencies,~$\nu_{\rm best} (n, l)$, and stellar parameters, such as mass, metallicity, age, etc., that describes the properties of the star but does not correctly model the frequencies due to the surface effect. A frequency-dependent correction,~$\nu_{\rm corr}$, needs to be added to the best-fitting model frequencies to eliminate the difference:
\begin{equation}
\label{equ:surfcorr}
\nu_{\rm obs} \simeq \nu_{\rm best} + \nu_{\rm corr}(\nu_{\rm best}).
\end{equation}

In practice, a grid of models is unlikely to include \change{the best-fitting model that describes the properties of the star. We homologously scaled} the mode frequencies of a closely fitting model by a factor $r$ \change{to} approximate \change{a better fitting} model \citep[see][]{kjeldsen08},
\begin{equation}
\label{equ:rscale}
\nu_{\rm best} \simeq r \nu_{\rm mod}.
\end{equation}
Scaling the mode frequencies by a factor $r$ changes $\Delta \nu$ by the same ratio, hence the density will change by a factor $r^2$. However, all other parameters, such as mass, effective temperature, initial metallicity cannot be easily tracked under an $r$-scale transformation. We assumed that other stellar parameters vary linearly by a small amount under this transformation. \change{This assumption includes the surface correction, that is, the frequency correction for the scaled model is a good estimate for the true frequency difference $\nu_{\rm corr}(\nu_{\rm best}) \simeq \nu_{\rm corr}(r \nu_{\rm mod})$.} Naturally, we made our analysis prefer a scale factor $r$ as close to unity as possible. 

\subsubsection{Power law correction}

We investigated three different types of surface correction from the literature. The first was proposed by \cite{kjeldsen08}, who used a single-term power-law fit to correct the mode frequencies. The authors originally made the correction as a function of the observed frequencies. However, we have chosen to use the frequencies of the best-fitting model, such that
\begin{equation}
\label{equ:kjeldcorr}
\nu_{\rm corr} = a \left( r \frac{\nu_{\rm mod}}{\nu_{\rm max}} \right)^b,
\end{equation}
which makes Eq.~\ref{equ:surfcorr} just a function of the model frequencies and is consistent with the other corrections we used. The frequency of maximum power,~$\numax$, was used to scale the frequencies. The exponent,~$b$, was originally fitted by \cite{kjeldsen08} using the discrepancy of observed and model frequencies in the Sun, and found to be $b \simeq 4.8$. However, there is no physical reason the solar-calibrated exponent should fit best for other stars. We expect it to be similar for stars with similar stellar properties but could potentially vary for other stars.

We considered two options when implementing the power-law surface correction. The first was to set $b$ to a constant value that would appropriately fit all stars in our sample, and the second was to let $b$ differ. The former was the method that we eventually used, with a constant of $b = 3$. The primary reason for this choice was that it is the same exponent used in two other surface corrections we tested, which will be introduced below. 

It can be argued that we were not faithful to the original \cite{kjeldsen08} correction. If the exponent was too close to one the correction would become unrealistically large with a scale factor $r$ far from unity. Alternative approaches we considered included making $b$ vary between stars by either making it a free parameter or find a relationship between $b$ and the stellar parameters. The former approach also strongly favours an exponent that is too close to one and was equally impractical. We briefly investigated the latter, which will be discussed further in the paper. Specifics on how the correction amount is dependent on the exponent in power-law method will be discussed in Section~\ref{sec:kjeldres}.  

To solve for $\nu_{\rm corr}$ we combine and rearrange Eq.~\ref{equ:surfcorr}~-~\ref{equ:kjeldcorr}. The observed frequencies can now be modeled with the following equation:
\begin{equation}
\label{equ:kjeldsen}
\nu_{\rm obs} \simeq r \nu_{\rm mod} + a \left( \frac{r \nu_{\rm mod}}{\nu_{\rm max}} \right)^b.
\end{equation}

\subsubsection{Cubic Correction}

\cite{ball14} introduced two new formulations of the surface correction\change{, inspired by \cite{gough90}}. The first considers the correction to be in the form of a frequency-cubed term divided by the mode inertia, referred to as the cubic correction,
\begin{equation}
\label{equ:cubiccorr}
\nu_{\rm corr} = c \left( \frac{\nu_{\rm best}}{\nu_{\rm max}} \right)^3/\mathcal{I}_{nl},
\end{equation}
where $\mathcal{I}_{nl}$ is the mode inertia for radial order $n$ and azimuthal degree $l$, and we solve for the coefficient $c$. The mode inertia is normalized at the radius of the star \citep{aerts10} and given by,
\begin{equation}
\label{equ:inertia}
\mathcal{I}_{nl} =  \frac{ 4 \pi \int^R_0 \left[ |\xi_r (r)|^2 + l(l+1)|\xi_h (r)|^2 \right] \rho r^2 dr}{\left[|\xi_r (R)|^2 + l(l+1)|\xi_h (R)|^2\right]}.
\end{equation}
Here, $\xi_r$ and $\xi_h$ are the radial and horizontal components of the displacement eigenvector, respectively. We calculated the mode inertia at $\numax$ ($\mathcal{I_{\numax}}$) by interpolating the mode inertia between the two adjacent modes near $\numax$. \change{Note that we are assuming that the modelled mode inertia is close to the true value, even after homology scaling.}


\begin{figure*}
\centering
 \includegraphics[width=2.0\columnwidth]{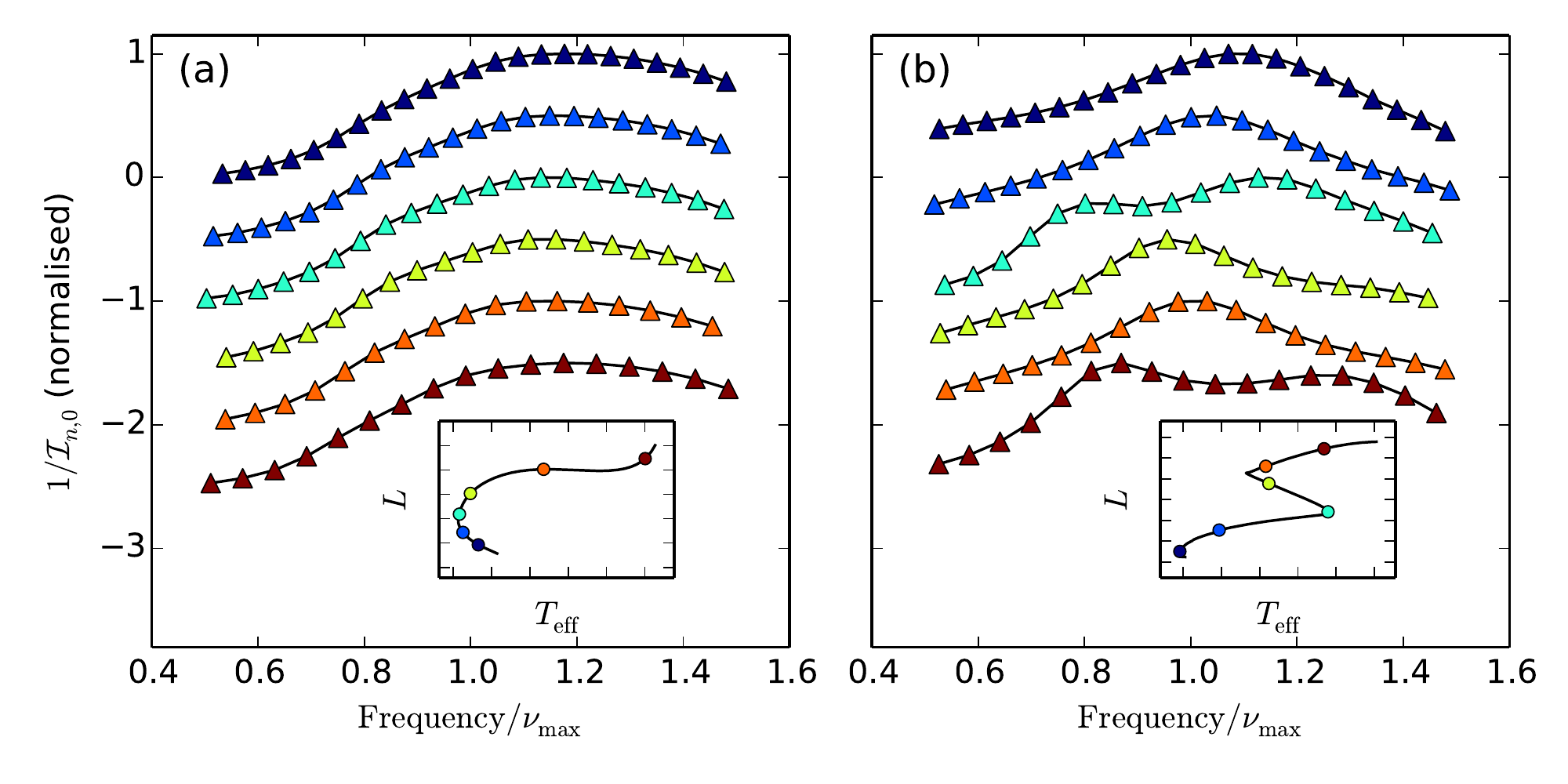}
 \caption{The inverse mode inertia for $l=0$ modes plotted against frequencies \change{normalized} by the scaling relation $\numax$ for two initial masses $M=1.0$ and $M=1.5$ (a and b respectively). Each model is normalized and offset by 0.5 for clarity.  Colours of the circle and triangle symbols in each panel represent the same model. Inset plot is the HR diagram of the model track (black line) and coloured symbols are the temperature and luminosity where each model was sampled.}
 \label{fig:figure10}
\end{figure*}
The profile of the inverse mode inertia over the course of main-sequence and early sub-giant evolution is shown in Fig.~\ref{fig:figure10} for a Sun-like stellar model ($M = 1.0 M_{\odot}$) and a higher-mass model ($M = 1.5 M_{\odot}$), which roughly bracket the mass range of the LEGACY sample. The mode inertia profile for the Sun-like star does not noticeably change, even as it approaches the sub-giant branch. However, the higher-mass star shows non-linear variation as it evolves. The double-hump feature in Fig.~\ref{fig:figure10}b before and after the end of main-sequence hook will affect the surface correction when mode inertia is included.


\subsubsection{Inverse-cubic Correction}

The second parameterization introduced by \cite{ball14} adds an inverse frequency term to the cubic term in Eq.~\ref{equ:cubiccorr}:,
\begin{equation}
\label{equ:ballcorr}
\nu_{\rm corr} = \left[ c_{-1} \left( \frac{\nu_{\rm best}}{\nu_{\rm max}} \right)^{-1} + c_{3} \left( \frac{\nu_{\rm best}}{\nu_{\rm max}} \right)^3 \right]/\mathcal{I}_{\nu},
\end{equation}
where we now solve for $c_{-1}$ and $c_3$, which are the inverse and cubic coefficients, respectively. Eq.~\ref{equ:ballcorr} is referred to as the inverse-cubic correction. This method has been shown to more accurately correct the mode frequencies in higher-mass main sequence stars using 3D hydrodynamic simulations \citep[e.g][]{ball16}. However, these corrections have not been tested on an ensemble of main sequence stars that sample a wide range of temperatures and masses as we set out to do here.


\subsubsection{Other corrections}

Along with the three surface corrections above, there are other approaches that attempt to correct the discrepancy in the models. We have not implemented them, but we note them here for completeness.
\begin{itemize}
\item Frequency difference ratios were proposed by \cite{roxburgh03, roxburgh13} to marginalise the effects of improper modelling of physics on the stellar surface layers. They are commonly used instead of absolute frequencies \citep{lebreton14,silva15}.
\item \cite{gruberbauer12} proposed a Bayesian method which includes a frequency offset parameter for each oscillation mode. They applied their method to both the Sun and a number of \kepler targets \citep[see][]{gruberbauerguenther13,gruberbauer13}. While the former proved successful, they did find biases towards stars with fewer low-frequency modes.
\item \cite{sonoi15} used a modified Lorentzian to fit the difference between standard 1D stellar models to 3D hydrodynamical simulations.
\end{itemize}
\section{Data and Models}
\label{sec:data}

\subsection{\it Kepler Data}
\label{sec:kepler}

\begin{figure}
 \includegraphics[width=\columnwidth]{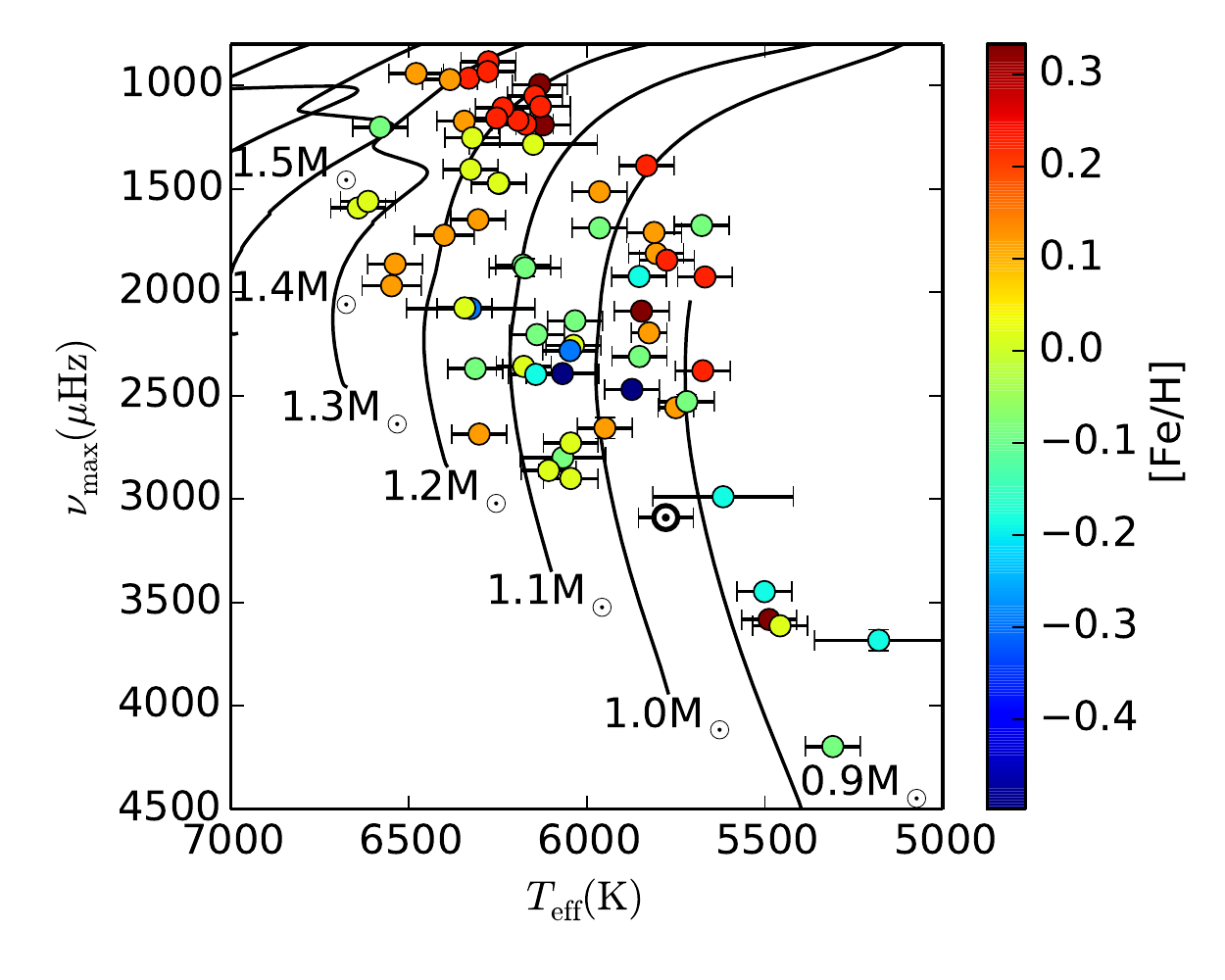}
 \caption{Modified Hertzsprung-Russell diagram of the LEGACY sample, with colour of the symbols corresponding to observed iron abundance. The lines are our stellar evolution models of differing mass at solar-metallicity calculated using the method outlined in \ref{sec:models}. The modelled $\numax$ was calculated using the asteroseismic scaling relations, Eq.~\ref{equ:numaxscale}.}
 \label{fig:figure2}
\end{figure}

We analysed the {\it Kepler} LEGACY sample \citep{lund17,silva17}, which consisted of 66 main-sequence \kepler stars where at least 12 months of short cadence data (one minute sampling) were available. Fig.~\ref{fig:figure2} shows the distribution of stars in our sample in the $\numax$-$T_{\rm eff}$ plane.

For each star in the sample, we used the frequency and corresponding uncertainties measured by \cite{lund17}. \change{It must be noted that their peak-bagging neglected mode asymmetry \citep[see][]{benomar18}, which contributes to the surface effects}. In general, temperatures and metallicities were adopted from the Stellar Parameters Classification (SPC) tool \change{\citep[see][]{buchhave12, buchhave15}}. For a small number of stars, temperatures and metallicities were from one of the following: \cite{ramirez09,huber13,casagrande14,chaplin14,pinsonneault12} \citep[see Table 1 from][for details]{lund17}. However, for two stars, KIC 9025370 and 9965715, we chose different effective temperatures than the ones stated by \cite{lund17}. \change{The quoted temperatures, which were $5270 \pm 180$K and $5860 \pm 180$K respectively, were} clear outliers in our initial results before we adopted alternative effective temperatures. The newly adopted temperatures were: $5617$K for KIC 9025370 from the \kepler Input Catalog, with the same uncertainty adopted by \cite{huber14} of $\pm 3.5$\%, and $6326 \pm 116$K for KIC 9965715 derived by \cite{molenda13}. 

\cite{lund17} included an analysis of the Sun using data from the VIRGO instrument \citep{froehlich09}. They reduced the solar time series using the same technique as the \kepler stars. Therefore, our sample consists of the 66 \kepler stars and also includes the Sun. We took the effective temperature of the Sun as if it were also taken from the SPC, therefore, the we assumed the uncertainty to be $\pm 77$K. 

\subsection{Stellar Models}
\label{sec:models}

We used the stellar evolution code MESA\footnote{\url{http://mesa.sourceforge.net/}} \citep[revision 9793,][]{paxton11,paxton13,paxton15} to calculate a grid of models. Unless otherwise stated, we used default options as described in the MESA documentation and source code. The models were parameterized by initial mass and metallicity,~$Z$. Mass was sampled between 0.84 and 1.64 solar masses with a spacing of 0.02 solar masses and an iron abundance between -0.6 to 0.4 dex with a spacing of 0.1 dex. This specific sampling was chosen because it would cover almost all the stars in our sample based on the results by \cite{lund17} and \cite{silva17}. The stars that were not contained on our grid were two lower-metallicity stars (KIC 7106245 and 8760414) and two lower-mass star (KIC 7970740 and 11772920). However, with the inclusion of a scale factor $r$ we aim to find an appropriately fitting model, even for stars falling slightly outside our model grid.

We adopted an enrichment law to calculate the initial helium abundance,~$Y$:
\begin{equation}
\label{equ:enrichment}
Y = 0.24 + 2Z.
\end{equation}
Mixing-length theory \citep{cox68} was used to describe convection, but set to a constant $\alpha_{\rm MLT} = 1.8$ in order to limit the dimensionality of the grid. Overshoot mixing was treated with the exponential decay formalism \citep{freytag96} with a value of $f_{\rm ov} = 0.01$. Our grid was intentionally not solar calibrated because we did not want a bias towards the more Sun-like stars, which were of less interest. We adopted a standard Eddington-grey atmosphere. OPAL opacities and chemical abundances used were derived by \cite{asplund09}. In general, we used the NACRE compilation \citep{angulo99} for thermonuclear reaction rates, but used the JINA \texttt{reaclib} database \citep{cyburt10} for $^{14}{\rm N}(p, \gamma)^{15}{\rm O}$, and \cite{kunz02} for $^{12}{\rm C}(\alpha, \gamma)^{16}{\rm O}$. Diffusion, gravitational settling, rotation, and mass loss were neglected. 

Models were evolved from pre-main-sequence until the large separation from the scaling relation reached the limit $\dnu = 40 \mu$Hz (smaller than for any star in our sample) or when the age exceeded 14 Gyr. To ensure no discontinuities in stellar structure, we enforced a maximum temperature change of $|\Delta T_{\rm eff}| < 20 {\rm K}$ between sequential models to avoid too large time steps. 

Adiabatic oscillation frequencies and mode-inertias were calculated using GYRE \citep[][]{townsend13}. The radial, dipole, and quadrupole modes were evaluated for models after the zero age main. Templates for our MESA and GYRE inlists are available online\footnote{\url{http://www.physics.usyd.edu.au/~dcom1502/inlists/}}.


\section{Method}
\label{sec:method}

For each star in our sample, we used the observed absolute mode frequencies and the iron abundance [Fe/H] to determine the probability for each model in our grid. We used a least-squares optimization routine to find the best-fitting scale factor $r$. The routine would search the one dimensional parameter space of $r$ and calculate the one or two coefficients of the empirically calculated surface correction using linear regression \citep[see][for details of calculation]{ball14}. The reduced $\chi^2$ statistic was calculated, defined by:
\begin{equation}
\label{equ:chi2}
\chi^2_\nu = \frac{1}{N_{\rm obs}}\sum_i^{N_{\rm obs}} \left( \frac{r\nu_{{\rm mod},i} - \nu_{{\rm obs},i}}{\sigma_{{\rm obs},i}} \right)^2,
\end{equation}
where $\sigma_{\rm obs}$ are the observed uncertainties and $N_{\rm obs}$ is the number of observed modes. The corresponding likelihood,
\begin{equation}
\label{equ:likelihood}
\mathcal{L}_\nu = \exp{\left(-\frac{1}{2} \chi_\nu^2\right)},
\end{equation}
defines the probability that the oscillation modes fit a given model. All available observed modes calculated by \cite{lund17} were used in our analysis.

With the introduction of the scale factor $r$, it was common to find multiple models with different initial parameters that provided a similarly good fit. The inclusion of iron abundance in the posterior calculation reduced the number of multiple solutions. Specifically, the ratio of iron-to-hydrogen for the stellar models was estimated from the initial metallicity of each model using the solar value 
\begin{equation}
\label{equ:feonh}
[{\rm Fe}/{\rm H}]_{\rm mod} = \log{(Z/X)} - \log{(Z/X)_{\rm sun}},
\end{equation}
where $(Z/X)_{\rm sun} = 0.0181$ \citep[from][]{asplund09}. The result from Eq.~\ref{equ:feonh} was used to calculate the metallicity likelihood,
\begin{equation}
\label{equ:metallike}
\mathcal{L}_{[{\rm Fe}/{\rm H}]} = \exp{\left(-\frac{1}{2} \frac{[{\rm Fe}/{\rm H}]_{\rm mod}-[{\rm Fe}/{\rm H}]_{\rm obs}}{\sigma_{[{\rm Fe}/{\rm H}]}}\right)},
\end{equation}
where $\sigma_{[{\rm Fe}/{\rm H}]}$ is the metallicity uncertainty from \cite{lund17}. The total likelihood was calculated by multiplying the two likelihoods (from Eq.~\ref{equ:likelihood}~and~\ref{equ:metallike}) together. We also only considered models with temperatures that were within three standard deviations from the observed values. That is, the temperature likelihood was one if $|T_{\rm eff, obs} - T_{\rm eff, mod}| \leq 3 \sigma_{T_{\rm eff}}$, else it was zero and the model was not considered.

It is important to note that using a scale factor $r$ changes the parameters of our model in a non-linear way (except mean density). Therefore, we included a Gaussian prior on our models where the probability of a model was greatest at $r=1$ with a standard deviation of 1\%. The width of the prior is an intentionally overestimated approximation to the relative frequency difference between sequential models of a typical oscillation mode around $\numax$.

The posterior was calculated for all models in our grid for each observed star. To determine the modelled stellar and surface correction parameters (and their uncertainties), the probability of the best-fitting model for each combination of initial mass and metallicity was used to weight the mean and standard deviation. A weighted average should average out any bias caused by the scale factor $r$. We included the weight of only one model for each evolutionary track to eliminate the bias caused by different steps that the stellar models take in parameter space. For example, the probability would be biased towards an evolutionary track that spent more time at the desired observed large separation, that is, where $\partial \nu/\partial t$ is small. Marginalising over time is not trivial because the time steps between sequential models were not equal, and biased towards more rapidly evolving stars. Additionally, we made the assumption that the probability distribution was symmetric before and after the best-fitting model for a given evolutionary track. Therefore, only considering a single model per track was a suitable approximation.

Finally, in some aspects of our analysis, we only considered the model with the greatest probability, in which the best-fitting model was used to compare the quality of fit between the surface corrections. We will mention below if a value is from the weighted average or the best-fitting model.



\section{Results}
\label{sec:results}


\subsection{Model results}
\label{sec:modres}

The weighted averages of the stellar parameters, and their uncertainties, are shown in Tables~A.\ref{tab:table_k}--A.\ref{tab:table_b}, along with the relative surface correction at $\numax$. It is difficult to independently verify our results because the {\it Kepler} data is the only way to estimate the parameters for most of the stars. Therefore, to test consistency, we considered the technique outlined by \cite{silva17}. They used the \basta pipeline \citep[see][]{silva15} as a reference to test the results against the other six pipelines in that paper. 
\begin{figure*}
 \includegraphics[width=2.\columnwidth]{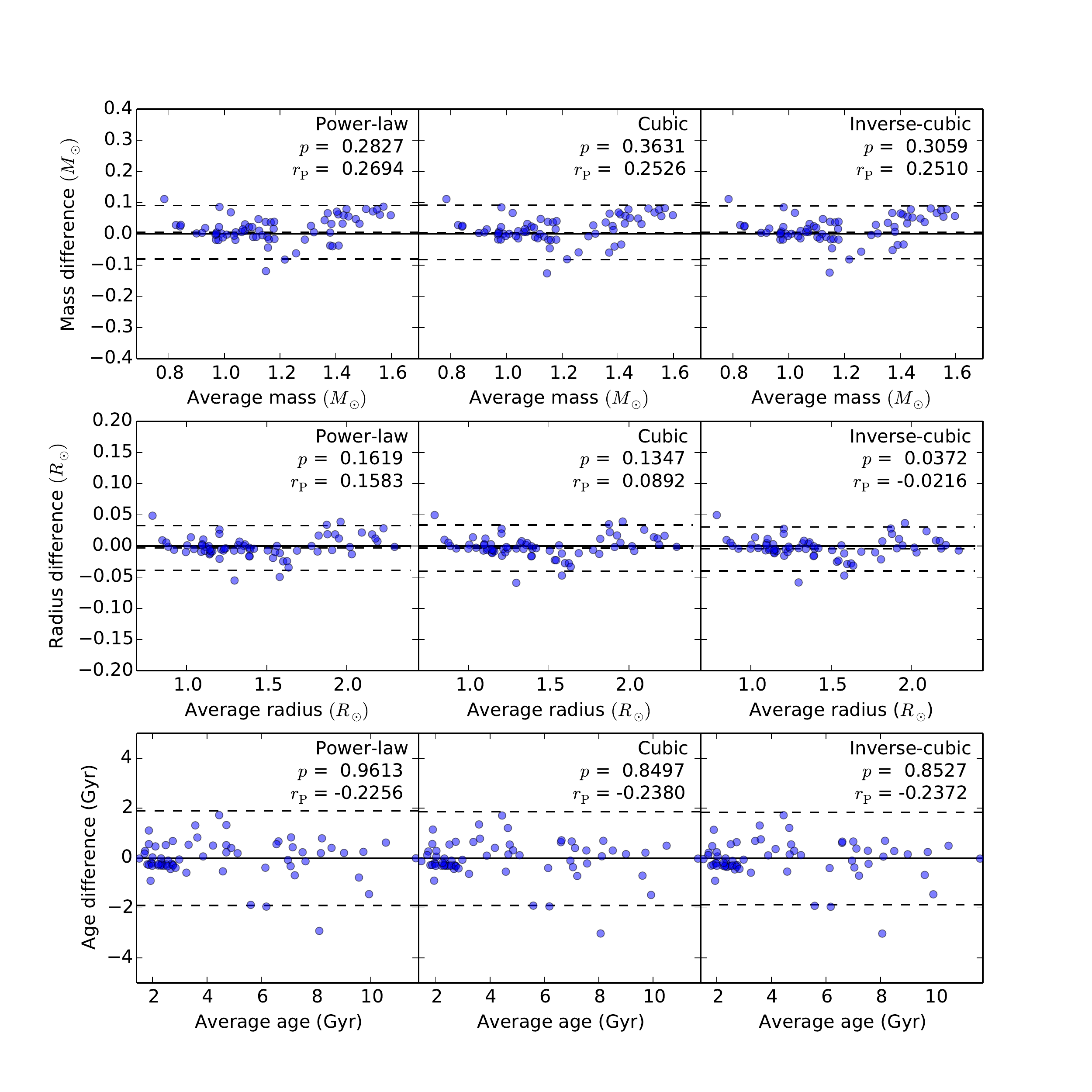}
 \caption{Plots of the absolute difference in mass, radius, and age between our pipeline and the \basta pipeline for all surface correction methods as a function of the average of our pipeline and the \basta pipeline. The solid lines mark where there is zero difference. The dashed lines correspond to the mean and twice the standard deviation of the differences weighted by their uncertainties. The $r$- and $p$-values are also shown (see text for details).}
 \label{fig:figure12}
\end{figure*}
Likewise, in Fig~\ref{fig:figure12} we plot the absolute differences between the results of the stellar parameters from \basta and our pipeline against the average of the two. The difference between each correction method and the \basta pipeline was greater than the difference across the three corrections. This strongly suggested the choice of correction method affects the resulting fitted stellar parameters much less than the effect from differences in the adopted physics of the stellar models.

We implemented the two quantitative diagnostics used by \cite{silva17} in their comparison: the Pearson product-moment correlation coefficient $r_{\rm P}$ (subscript added to avoid confusion with the homology scale factor $r$) and the p-value of a one-sample t-test of the weighted mean of the differences. Both values are shown in Fig~\ref{fig:figure12}. Based on the sample size, a correlation coefficient of $|r_{\rm P}| > 0.25$ from a linear regression analysis indicates a significant \change{difference} at a level of 5\%. While the mass does show a slight \change{difference}, the radius and age parameters do not. However, \change{our calculated radius was consistently greater than the \basta results for the larger stars}. This was likely due to the adoption of a constant mixing length for our models. Calibrations of the mixing length parameter using 3D hydrodynamical simulations \citep[e.g.][]{trampedach14,magic15} show that the mixing length decreases as surface gravity decreases or effective temperature increases. Therefore, our mixing length is clearly overestimated for the higher-mass stars in our grid.


We also considered the \change{difference in the means of the results (weighted by the uncertainties) between the two pipelines}. Following \cite{silva17}, we rejected the null hypothesis ${\rm H}_0$ if the $p$-value was less than 0.05. The radius calculated using the inverse-cubic correction was the only parameter that showed \change{a significant difference}. Other \change{discrepancies} will be discussed in the in Section~\ref{sec:short}. We did not consider alternative methods of comparison such as the parallaxes or interferometric radii, which were only available for a small subset of our sample, because we wanted to focus on the general trends of all stars in the sample (obtaining the stellar parameters was not the primary goal of this paper).

\begin{figure*}
 \includegraphics[width=2\columnwidth]{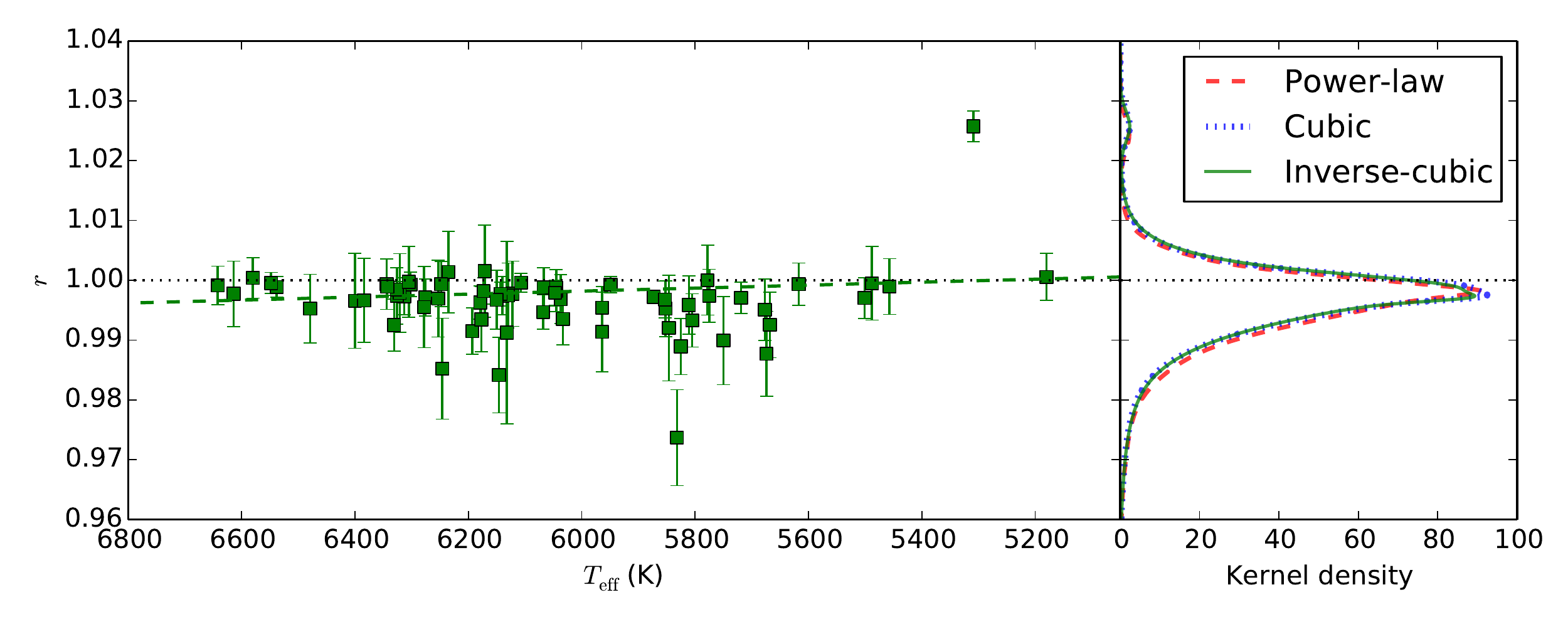}
 \caption{The left panel shows the weighted average of the calculated scale factor $r$ for the inverse-cubic method against effective temperature. Error bars show one-sigma uncertainties. The green dashed line shows the weighted linear fit to the data. The right panel is the kernel density estimate of the $r$-values for all three correction methods.}
 \label{fig:figure11}
\end{figure*}
For each star, the mean and standard deviation of the scale factor $r$ was calculated using posterior probability of the models as weights. We examine the results of the homology scaling in Fig.~\ref{fig:figure11}. Compared to \cite{kjeldsen08}, where they allowed the frequencies to be scaled by over $11\%$ for a model of $\beta$~Hyi, we were more cautious when allowing the frequencies to be scaled away from unity. The distribution of calculated $r$-values show a clear fixed bias slightly towards $r < 1$. No significant proportional bias ($|r_{P}| < 0.25$) can be seen in Fig.~\ref{fig:figure11}, but the slope of the weighted linear fit is still dominated by the star in the top-right of the plot. In fact, the slope is only negative because of that star. 

The clear 'outlier' star in the left panel of Fig.~\ref{fig:figure11} is KIC 7970740, a low-mass star that was known to have a mass outside of our grid range. Hence, our calculated stellar parameters shown in Tables~\ref{tab:table_k}-\ref{tab:table_b}, particularly mass, are less accurate for this star because the homology scaling was used to extrapolate our grid. It was expected that modelled mode frequencies would need to be increased to match the observed frequencies. The star with lowest $r$ value, KIC 6933899, is not a significant outlier and the stellar and surface correction parameters are in agreement with the rest of the sample. However, this star has a lower scale factor $r$ because it probably has an inaccurate helium abundance due to the correlated residuals in the mode frequencies (black symbols in Fig.\ref{fig:figure3}). The reason for this will be discussed in Section~\ref{sec:short}.

The average $r$-value for each star was within the one-sigma uncertainty for all three correction methods resulting in three nearly identical kernel densities. This reinforces the fact that the the stellar parameters from our pipeline are mostly independent of the chosen correction method. Overall, we are satisfied with the implementation of the scale factor $r$, and its ability to produce better fitting models.





\subsection{Surface correction results}
\label{sec:surfres}

\begin{figure*}
\centering
 \includegraphics[width=2.1\columnwidth]{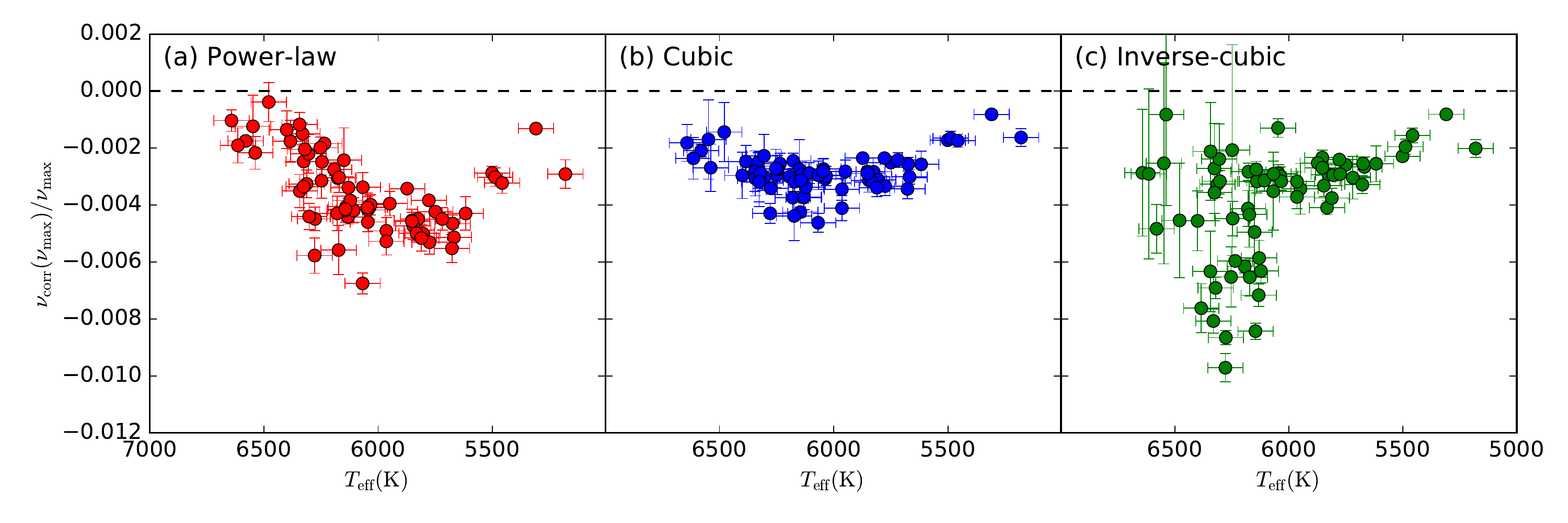}
 \caption{Relative surface corrections, and their uncertainties, at $\numax$ as a function of effective temperature for the LEGACY sample: (a) using the power-law correction, (b) the cubic method, and (c) the inverse-cubic term method. Temperatures and uncertainties are from \protect\cite{lund17}.}
 \label{fig:figure7b}
\end{figure*}

\begin{figure*}
\centering
 \includegraphics[width=2.1\columnwidth]{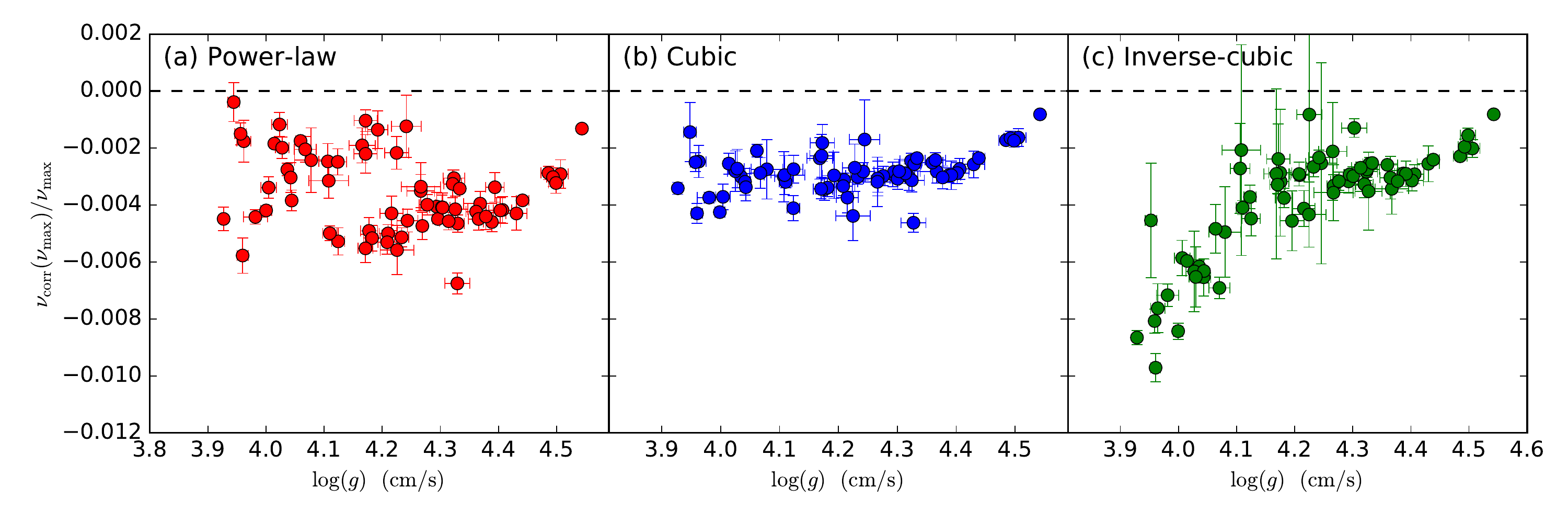}
 \caption{Same as Fig.~\ref{fig:figure7a} but as a function of surface gravity. Surface gravities and uncertainties were calculated using the weighted mean from our models.}
 \label{fig:figure7a}
\end{figure*}

We present the ensemble analysis of the three surface corrections in Figs.~\ref{fig:figure7b} and \ref{fig:figure7a}. The results confirm that the surface correction is always negative. This is encouraging because there were no priors on the surface correction coefficients. 


We expected to observe trends between the surface correction and those stellar parameters that probe the outer layers of the star. Figs.~\ref{fig:figure7b} and \ref{fig:figure7a} show the relative surface correction at $\numax${\color{blue}, $\nu_{\rm corr}(\numax)/\numax$,} against effective temperature and surface gravity, respectively. These two parameters were chosen specifically because they are the two independent variables in the $\numax$ scaling relation,~Eq.~\ref{equ:numaxscale}. Additionally, \cite{trampedach17} showed, using 3D stellar surface convection simulations, a relationship between the relative surface correction and these two stellar parameters for main-sequence stars. Some of these figures display a clear trend and could imply a causal relationship. They will be discussed in their corresponding subsections below. Throughout our discussion we will show the details of the surface correction for the two stars first shown in Fig.~\ref{fig:figure1}, KIC 4914923 and KIC 12317678. Note that we will be using the frequencies from the best-fitting scaled model (the model with the greatest total likelihood) in our comparison of these two stars. These two stars would traditionally be classified as simple and F-like, respectively \citep[see][]{appourchaux12}.

\subsubsection{Power-law results}
\label{sec:kjeldres}

\begin{figure*}
 \includegraphics[width=2.\columnwidth]{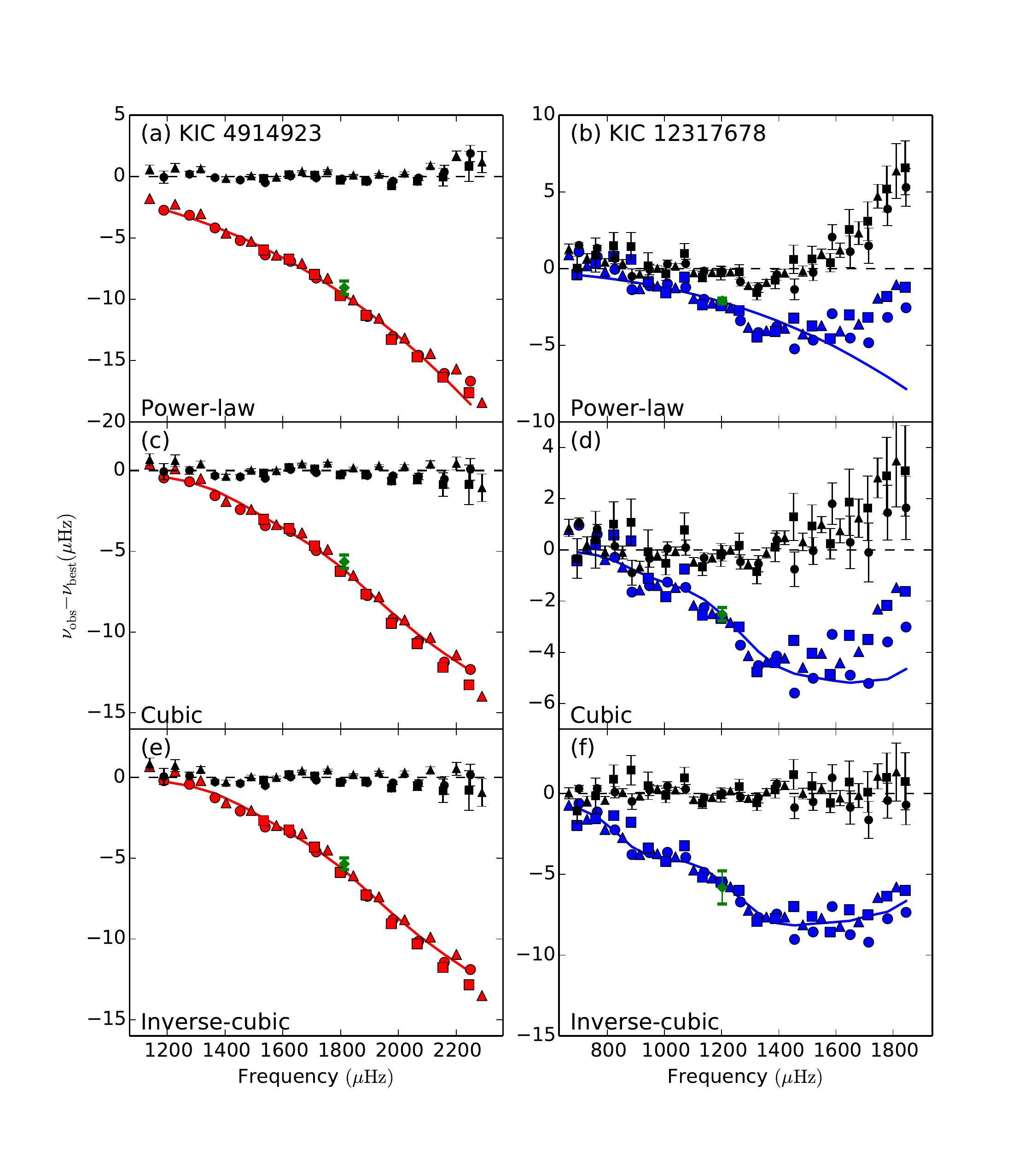}
 \caption{Difference between the observed and modelled frequencies for two stars in the LEGACY sample, KIC 4914923 (left) and KIC 12317678 (right), before (red/blue symbols, respectively) and after (black symbols) each surface correction. The different symbols represent the different angular degrees and are the same as in Fig.~\ref{fig:figure1}. \change{The coloured lines represent the calculated correction $\nu_{\rm corr}$ as a function of frequency}. The green diamonds mark the weighted average correction at $\numax$ and its uncertainty.}
 \label{fig:figure3}
\end{figure*}

The results for the power-law correction in Fig.\ref{fig:figure7a}a show a strong correlation between temperature and the amount of relative surface correction at $\numax$. Specifically, we observed \change{the} greatest correction at approximately solar temperature, with the correction decreasing for hotter stars. Interestingly, the surface correction appears to decrease for stars cooler than the Sun, although the sample size is small. We are not able to suggest a physical reason for observing a maximum in the amount of surface correction at solar temperature. An example of a hotter star with negligible surface correction, KIC 12317678, is shown in Fig.~\ref{fig:figure3}b, which agrees with the analysis of the F star Procyon by \cite{bedding10}, where they fitted a model with no surface correction to the observed frequencies using a particular mode identification scenario. No trend cannot be seen as a function of surface gravity, shown in Fig.\ref{fig:figure7a}a. 


For the most Sun-like stars in our sample, the results from our pipeline do not agree with the work done by \cite{kjeldsen08}. The magnitude of our correction was greater than their analysis between the Sun's mode frequencies and Model S \citep{cd96}. Additionally, we notice an upturn in the residuals towards higher frequencies when using the power-law method (see Fig.~\ref{fig:figure3}a for an example of a Sun-like star). This feature is present in the results of most stars in our sample, and can be seen in our hotter star example shown in Fig.~\ref{fig:figure3}b. In our analysis, decreasing the exponent in the power-law correction star improved the fit, but also increased the magnitude of the correction to an implausible amount. This contradicts the analysis by \cite{kjeldsen08}, who found a greater value of $b \simeq 4.8$ to be the most suitable. Presumably, the differing physics between our grid of models and Model S can explain the difference between the corrections amounts. In Section~\ref{sec:kjeldrev} we will revisit the power-law correction and use the results for the cubic correction to estimate new and varying exponents to use for a power-law correction.






\subsubsection{Cubic results}
\label{sec:cubicres}


The addition of mode inertia in the surface correction makes a number of subtle yet important changes to the correction, without a significant change in the calculated stellar parameters. The inclusion of the mode inertia improved the likelihood ($\mathcal{L_{\rm Cubic}}/\mathcal{L_{\rm P-L}} > 1$) for all stars despite having the same number of free parameters as the pure power-law method. The relative surface correction at $\numax$ as a function of both effective temperature and $\log{(g)}$, shown in Figs.~\ref{fig:figure7b}b~and~\ref{fig:figure7a}b, shows a flatter trend than the power-law equivalent. In general, the magnitude of correction increases for the hotter stars, and decreases for the cooler stars compared to the power-law correction. Implementing the mode inertia into the correction is approximately equivalent to decreasing the exponent for the hotter stars and increasing it for the cooler stars, but with the advantage of also improving the overall fit to the mode frequencies.

Additionally, the profile of the mode inertia (see Fig.~\ref{fig:figure10}) introduces non-linearity to the surface correction, which is more pronounced in hotter stars. This reduces the upturn we saw with the power-law correction for the cooler stars. Finally, we observe a decrease in correction ($\nu_{\rm corr}$) at higher frequencies in the hotter star, shown in Fig.~\ref{fig:figure3}d. However, the residuals still show a slight, but reduced, upturn.

\subsubsection{Power-law revisited}
\label{sec:kjeldrev}

We will now take a detour to investigate a power-law method with a varied exponent. We attempted this in two ways: making the exponent a free parameter in our fit, or using the results of the cubic correction to infer an exponent and applying it to the power-law correction, while omitting the mode-inertia. Using a free parameter may seem like a natural choice, however, in conjunction with the homology scaling it produced unphysical model choices for our sample of stars. That is, the exponent would tend to make the correction linear, $b \simeq 1$, giving extremely large frequency corrections with a scale factor $r$ that would stray far from unity, despite the implementation of the $r$-restricting prior. The stellar parameters of the chosen model for a free parameter fit were outliers compared to the other correction methods.

\begin{figure}
\centering
 \includegraphics[width=1.0\columnwidth]{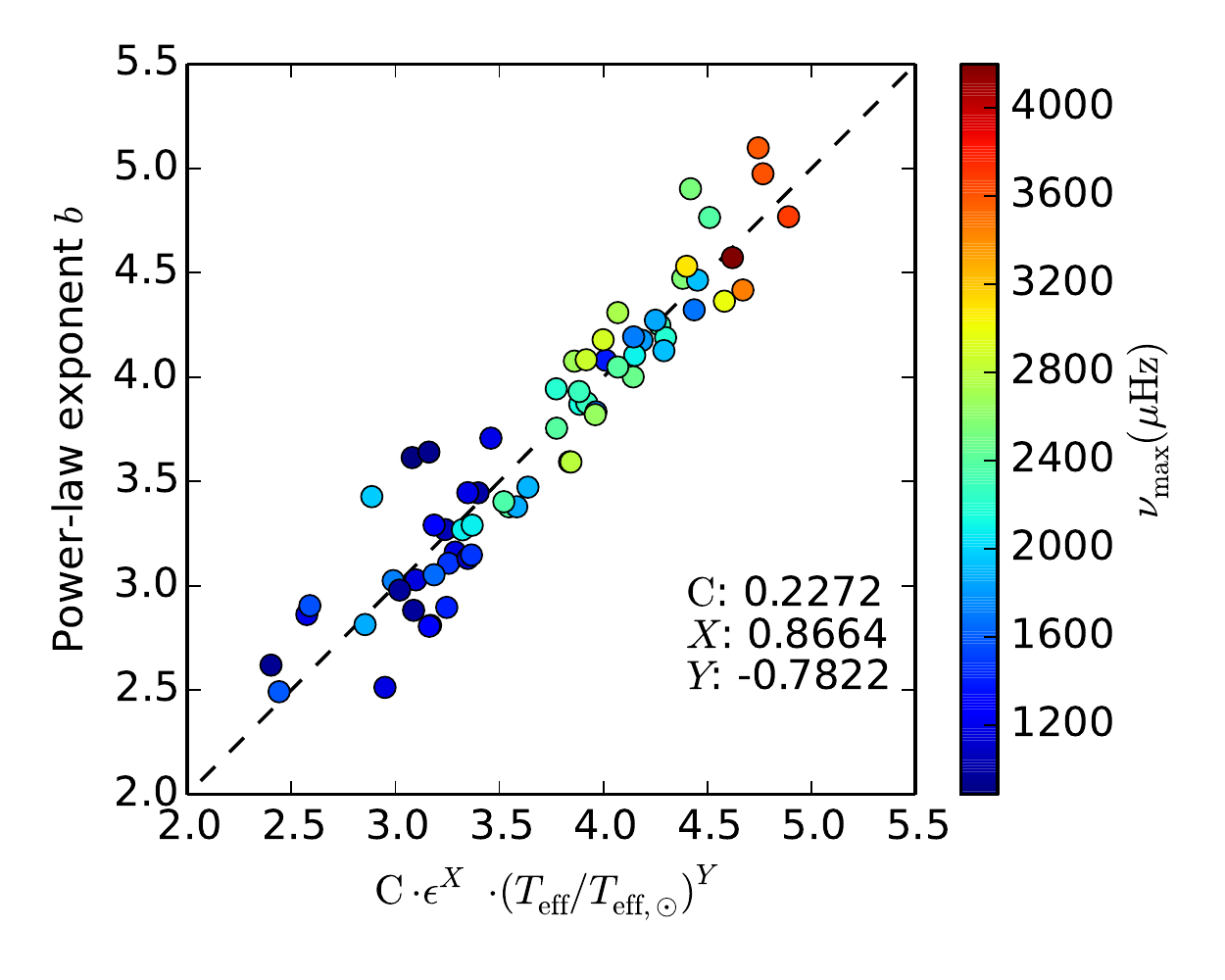}
 \caption{Results of fitting a power-law to the cubic frequency corrections $\nu_{\rm corr}$ as a function of an arbitrary scaling relation. See text for details.}
 \label{fig:figure14}
\end{figure}

The second method we attempted required fitting a two-parameter power law, like the functional form of Eq.~\ref{equ:kjeldcorr}, to the cubic correction and then to extract the exponent for each star. The resulting exponents ranged between values we would expect for high- and low-mass stars. We noticed a strong relationship between the exponent and other stellar parameters. Therefore, instead of directly using the exponents we extracted from the cubic correction in the power-law correction, we constructed a function based on the surface stellar parameters to estimate the exponent, similar to what \cite{sonoi15} did for the parameters in their Lorentzian correction method. We found that a scaling relation, as a function of effective temperature and $\epsilon$ from Eq.~\ref{equ:asym}, best described the fitted exponents. A plot of the exponents as a function of the arbitrary scaling relation, along with the scaling parameters, is shown in Fig.~\ref{fig:figure14}.

The fit to the cubic correction suggests that the power-law exponent is strongly related to the stellar parameters. In general, a higher exponent was found to fit better for the cooler stars in the sample. This contradicts what we found in Section~\ref{sec:kjeldres}, but agrees with the value of the Sun found by \cite{kjeldsen08}. We used this relation as input for the exponent of a power-law without mode inertia. The magnitude of the correction only changed slight improvement for the cooler stars in the sample and no significant change for the hotter stars compared to the pure power-law correction method. This was the result we expected, however, the quality of the fit did not improve compared to the constant exponent power-law we originally implemented. This analysis further suggests that non-linearity of the mode inertia is required to get appropriate fits for all stars in our sample. 

\subsubsection{Inverse-cubic results}
\label{sec:combinedres}

\begin{figure}
\centering
 \includegraphics[width=\columnwidth]{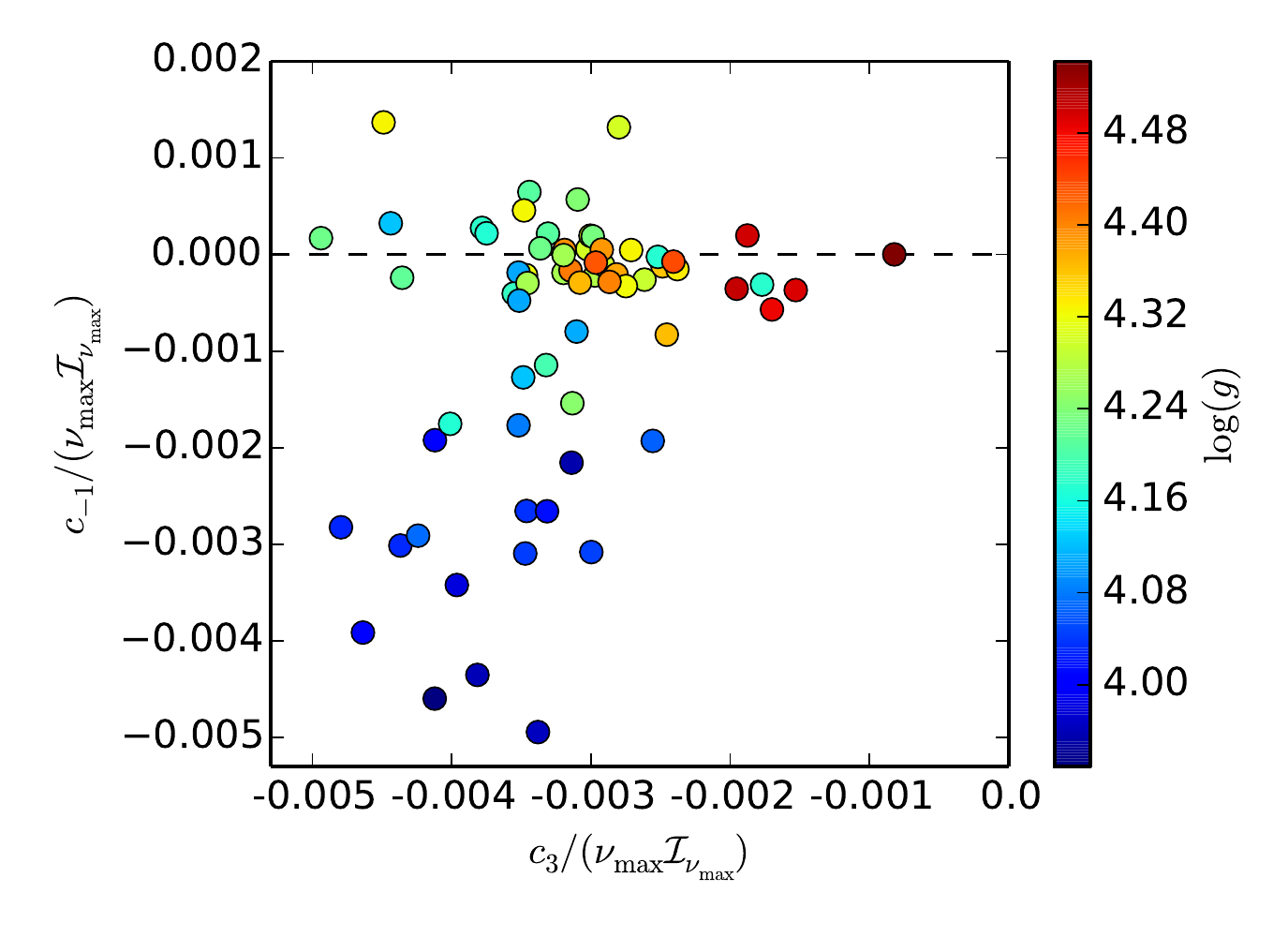}
 \caption{Distribution of the coefficients of the inverse-cubic surface correction for the LEGACY sample taken from the best-fitting model. The colour represents the surface gravity for the chosen model, calculated using the weighted average of the models (see text).}
 \label{fig:figure8}
\end{figure}

Finally, the addition of the inverse term produces the most dramatic changes of all three corrections. Fig.~\ref{fig:figure8} shows that the contribution of the inverse coefficient term is most noticeable in stars with lower surface gravity. For the cooler stars, there was little change compared to the cubic correction for stars with temperatures up to $6000{\rm K}$. This can be seen in our cool star example, Fig.~\ref{fig:figure3}e, which is indistinguishable from the cubic example in Fig.~\ref{fig:figure3}c.

The ensemble of stars show a relative surface correction that overall increases with temperature, shown in Fig.~\ref{fig:figure7b}c, the opposite of what we observed using the power-law correction. The divergence in this plot at high temperatures (hotter than $6000{\rm K}$) implies there are two groups of F stars which can be separated at approximately $\nu_{\rm corr}/\numax=-0.006$. We found that the differences in surface gravity was enough to distinguish between the two groups. Therefore, the relative surface correction correlated better with the surface gravity for the inverse-cubic method, shown in Fig.~\ref{fig:figure7a}c. 

\begin{figure}
 \includegraphics[width=\columnwidth]{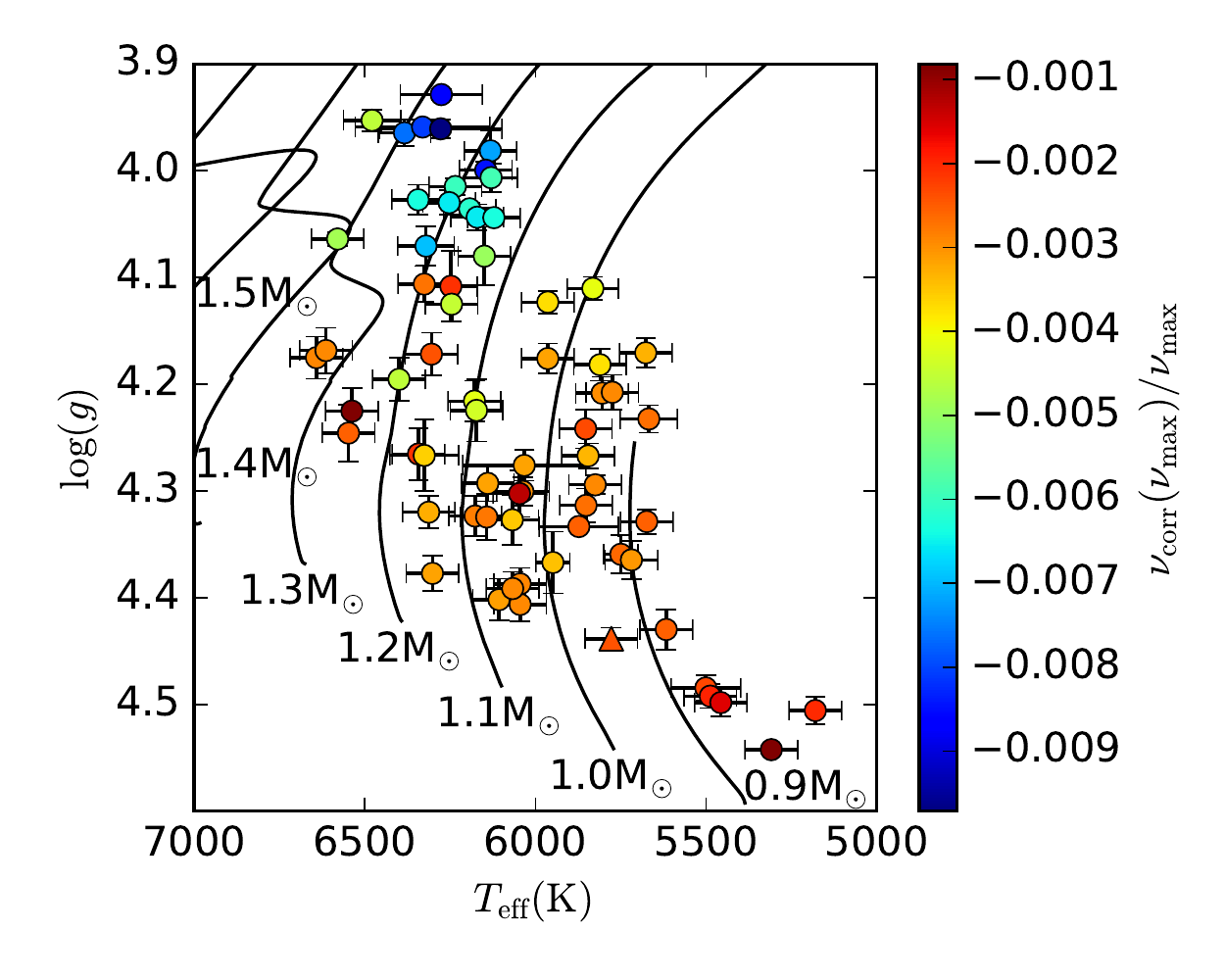}
 \caption{Hertzsprung-Russell diagram of our sample, with colour of the symbols corresponding the relative surface correction at $\numax$ \change{for the inverse-cubic results}. The triangle symbol represents the Sun. The black lines are our stellar evolution models of differing mass at solar-metallicity calculated using the method outlined in \ref{sec:models}.}
 \label{fig:figure22}
\end{figure}

Combining the results from Figs.~\ref{fig:figure7b}c and \ref{fig:figure7a}c we can visualise how the surface correction varies with age. Consider any set of observed stars in Fig.~\ref{fig:figure22} which lie near the same model track (black lines). In general, the relative surface correction increases for the more evolved stars on any given track. This agrees with the predicted surface correction from 3D hydrodynamical simulations \citep[see][]{trampedach17}, but only for the inverse-cubic correction.

\begin{figure}
\centering
 \includegraphics[width=\columnwidth]{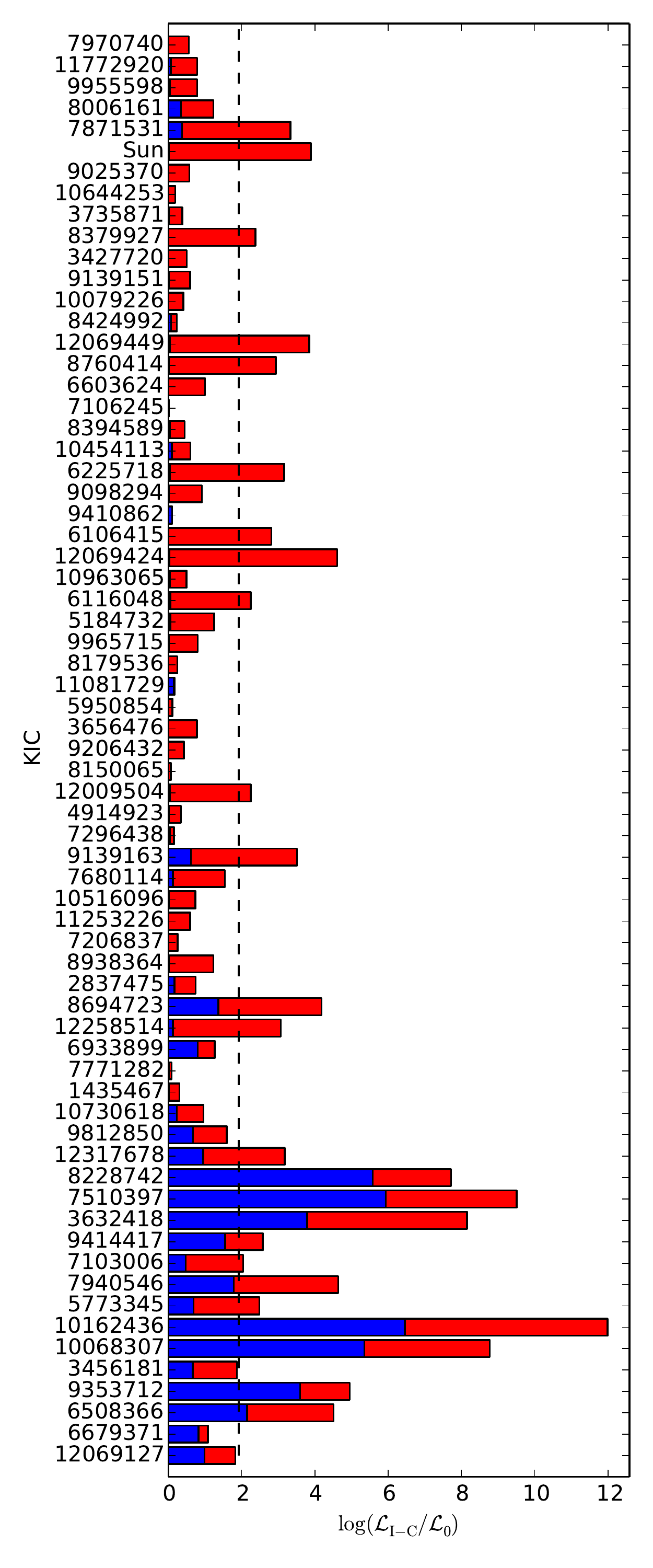}
 \caption{Logarithmic likelihood ratio between inverse-cubic method and the power-law (red) or cubic (blue) method. The dashed line represents the difference in logarithmic likelihood required such that likelihood ratio test is significant and $p<0.05$, due to additional degree of freedom in the inverse-cubic method. The y-axis is ordered by the surface gravity from the results of the inverse-cubic method (bottom is low surface gravity).}
 \label{fig:figure9}
\end{figure}

Fig.~\ref{fig:figure3}f shows an example of how the inverse-cubic term performs the best for the hotter stars, particularly by greatly improving the fit at high frequencies. As expected, the likelihoods of the inverse-cubic correction models are exclusively better than the other two methods for all stars, shown in Fig.~\ref{fig:figure9}. However, when we consider the extra degree of freedom, the goodness of fit of the inverse-cubic method is only \change{substantially better} for 25 stars compared to the power-law method and 7 stars for the cubic method, at a critical value of $p<0.05$. 


\subsection{Discussion}
\label{sec:short}

In general, more massive and evolved stars in our sample were more likely to have larger uncertainties due to the degeneracies around the end of main-sequence 'hook'. The model tracks in the top-left of Figs.~\ref{fig:figure2} and \ref{fig:figure22} show how a star with a given metallicity, temperature, and $\numax$ can be described by multiple masses. It was difficult to refine the fit for stars in this regime. The introduction of the scale factor $r$ and then the additional free parameter in the inverse-cubic correction allowed relatively good fits to a wider range of models. For a number stars when using the inverse-cubic correction, this greatly increased the uncertainties of the stellar and correction parameters, and can clearly be seen for a number of stars in Figs.~\ref{fig:figure7b}c and \ref{fig:figure7a}c. This was the only downside the two-term fit had over the other methods we tried. Stronger constraints on the stellar parameters would reduce the ambiguity. However, this would have required another independent method of measurement, such as stellar radii from interferometry \change{\citep[e.g.][]{white13}}.


The residuals, $\nu_{\rm obs} - (\change{r \nu_{\rm mod}} + \nu_{\rm corr})$, for a number of stars in our results showed a clear correlated scatter due to the assumption we made about the helium abundance in our models. For these stars, the enrichment law, Eq.~\ref{equ:enrichment}, was not a good estimation of the stellar composition. The use of the metallicity likelihood, Eq.~\ref{equ:metallike}, ensured that the modelled iron abundance was fit better to the observations than the helium abundance. Oscillation mode frequencies in solar-like stars are affected by the helium abundance which, in conjunction with the effective temperature, determines the depth of helium ionisation zone \citep[see][]{verma14b, verma17}. This leads to acoustic glitches that appear as sinusoidal variations in the asymptotic relation, e.g. Fig~\ref{fig:figure1}b. Fitting modelled frequencies to this curvature, for all angular degrees, would require an accurate composition for all elements in the model. Our model grid cannot easily remedy this problem without increasing the number of dimensions.


Our definition of the surface correction at $\numax$ does not fully capture the difference between the observed and model pulsation frequencies for cases where the correction does not monotonically decrease, e.g. Fig.~\ref{fig:figure3}f. However, we found no scalar that could better characterize the profile of the surface correction. While our analysis shows a strong correlation between the surface gravity and the relative surface correction at $\numax$, a better metric should be found that describes the non-linearity of these corrections.

\section{Conclusions}
\label{sec:conc}

We created a pipeline that was able to fit observed pulsation frequencies to a model grid using an empirically calculated surface correction. The model grid was created using MESA, parameterized by mass, metallicity, and age, with pulsation frequencies calculated using GYRE. The pipeline was used to \change{determine} the stellar and surface correction parameters for 66 observed main-sequence solar-like oscillators stars from \kepler as well as the Sun. \change{The resulting stellar parameters for all three surface correction methods agreed well with an independent pipeline. However, the results for each surface correction from our pipeline were much more similar to each other than the results from the \basta pipeline. This suggested that choice of adopted physics has a greater impact than the choice of surface correction}. Applying a scale factor $r$ to the cubic and inverse-cubic surface corrections allowed for easy interpolation between models and improved the robustness of our fits.

We found that including mode inertia in the surface correction always improved the frequency fit to the models without including another free parameter. The inverse term greatly improved the fit to the models for the more evolved stars without much change to the Sun-like stars. Therefore, the cubic term still dominated the correction for the Sun-like stars while the contribution of the inverse term becomes much more significant for older stars. This resulted in a larger relative correction for lower surface gravity stars, when using the inverse-cubic method. This agrees with the results found using 3D hydrodynamical simulations.

However, including another free parameter occasionally caused over-fitting for a small number of stars, greatly increasing the uncertainty of the surface correction for a small number of stars in our sample. Despite this, the inverse-cubic correction best describes the discrepancy between observed and model pulsation frequencies for all main-sequence stars. The surface correction can be used to constrain a stellar model because we have shown the correction correlates with the fundamental stellar parameters. We recommend using the inverse-cubic method to empirically correct for near surface effects for 1D stellar evolution models.

\section*{Acknowledgements}
\label{sec:ack}

This research was supported by the Australian Research Council. Funding for this project has been provided by the Group of Eight Australia-Germany Joint Research Cooperation Scheme. Funding for the Stellar Astrophysics Centre is provided by the Danish National Research Foundation (grant agreement no.: DNRF106). The research is supported by the ASTERISK project (ASTERoseismic Investigations with SONG and Kepler) funded by the European Research Council (grant agreement no.: 267864). WHB acknowledges support from the UK Science and Technology Facilities Council (STFC). TRW acknowledges the support of the Villum Foundation (research grant 10118). DH acknowledges support by the Australian Research Council's Discovery Projects funding scheme (project number DE140101364) and support by the National Aeronautics and Space Administration under Grant NNX14AB92G issued through the Kepler Participating Scientist Program.

\bibliographystyle{mnras}
\bibliography{ref}

\begin{thebibliography}{}
\makeatletter
\relax
\def\mn@urlcharsother{\let\do\@makeother \do\$\do\&\do\#\do\^\do\_\do\%\do\~}
\def\mn@doi{\begingroup\mn@urlcharsother \@ifnextchar [ {\mn@doi@}
  {\mn@doi@[]}}
\def\mn@doi@[#1]#2{\def\@tempa{#1}\ifx\@tempa\@empty \href
  {http://dx.doi.org/#2} {doi:#2}\else \href {http://dx.doi.org/#2} {#1}\fi
  \endgroup}
\def\mn@eprint#1#2{\mn@eprint@#1:#2::\@nil}
\def\mn@eprint@arXiv#1{\href {http://arxiv.org/abs/#1} {{\tt arXiv:#1}}}
\def\mn@eprint@dblp#1{\href {http://dblp.uni-trier.de/rec/bibtex/#1.xml}
  {dblp:#1}}
\def\mn@eprint@#1:#2:#3:#4\@nil{\def\@tempa {#1}\def\@tempb {#2}\def\@tempc
  {#3}\ifx \@tempc \@empty \let \@tempc \@tempb \let \@tempb \@tempa \fi \ifx
  \@tempb \@empty \def\@tempb {arXiv}\fi \@ifundefined
  {mn@eprint@\@tempb}{\@tempb:\@tempc}{\expandafter \expandafter \csname
  mn@eprint@\@tempb\endcsname \expandafter{\@tempc}}}

\bibitem[\protect\citeauthoryear{{Aerts}, {Christensen-Dalsgaard}  \&
  {Kurtz}}{{Aerts} et~al.}{2010}]{aerts10}
{Aerts} C.,  {Christensen-Dalsgaard} J.,   {Kurtz} D.~W.,  2010,
  {Asteroseismology}

\bibitem[\protect\citeauthoryear{{Angulo} et~al.,}{{Angulo}
  et~al.}{1999}]{angulo99}
{Angulo} C.,  et~al., 1999, \mn@doi [Nuclear Physics A]
  {10.1016/S0375-9474(99)00030-5}, \href
  {http://adsabs.harvard.edu/abs/1999NuPhA.656....3A} {656, 3}

\bibitem[\protect\citeauthoryear{{Appourchaux} et~al.,}{{Appourchaux}
  et~al.}{2012}]{appourchaux12}
{Appourchaux} T.,  et~al., 2012, \mn@doi [\aap] {10.1051/0004-6361/201218948},
  \href {http://adsabs.harvard.edu/abs/2012A%26A...543A..54A} {543, A54}

\bibitem[\protect\citeauthoryear{{Asplund}, {Grevesse}, {Sauval}  \&
  {Scott}}{{Asplund} et~al.}{2009}]{asplund09}
{Asplund} M.,  {Grevesse} N.,  {Sauval} A.~J.,   {Scott} P.,  2009, \mn@doi
  [\araa] {10.1146/annurev.astro.46.060407.145222}, \href
  {http://adsabs.harvard.edu/abs/2009ARA%26A..47..481A} {47, 481}

\bibitem[\protect\citeauthoryear{{Ball} \& {Gizon}}{{Ball} \&
  {Gizon}}{2014}]{ball14}
{Ball} W.~H.,  {Gizon} L.,  2014, \mn@doi [\aap] {10.1051/0004-6361/201424325},
  \href {http://adsabs.harvard.edu/abs/2014A%26A...568A.123B} {568, A123}

\bibitem[\protect\citeauthoryear{{Ball} \& {Gizon}}{{Ball} \&
  {Gizon}}{2017}]{ball17}
{Ball} W.~H.,  {Gizon} L.,  2017, \mn@doi [\aap] {10.1051/0004-6361/201630260},
  \href {http://adsabs.harvard.edu/abs/2017A%26A...600A.128B} {600, A128}

\bibitem[\protect\citeauthoryear{{Ball}, {Beeck}, {Cameron}  \& {Gizon}}{{Ball}
  et~al.}{2016}]{ball16}
{Ball} W.~H.,  {Beeck} B.,  {Cameron} R.~H.,   {Gizon} L.,  2016, \mn@doi
  [\aap] {10.1051/0004-6361/201628300}, \href
  {http://adsabs.harvard.edu/abs/2016A%26A...592A.159B} {592, A159}

\bibitem[\protect\citeauthoryear{{Ball}, {Theme{\ss}l}  \& {Hekker}}{{Ball}
  et~al.}{2018}]{ball18}
{Ball} W.~H.,  {Theme{\ss}l} N.,   {Hekker} S.,  2018, preprint, \href
  {http://adsabs.harvard.edu/abs/2018arXiv180411153B} {} (\mn@eprint {arXiv}
  {1804.11153})

\bibitem[\protect\citeauthoryear{{Bedding} et~al.,}{{Bedding}
  et~al.}{2010}]{bedding10}
{Bedding} T.~R.,  et~al., 2010, \mn@doi [\apj] {10.1088/0004-637X/713/2/935},
  \href {http://adsabs.harvard.edu/abs/2010ApJ...713..935B} {713, 935}

\bibitem[\protect\citeauthoryear{{Beeck}, {Cameron}, {Reiners}  \&
  {Sch{\"u}ssler}}{{Beeck} et~al.}{2013}]{beeck13}
{Beeck} B.,  {Cameron} R.~H.,  {Reiners} A.,   {Sch{\"u}ssler} M.,  2013,
  \mn@doi [\aap] {10.1051/0004-6361/201321343}, \href
  {http://adsabs.harvard.edu/abs/2013A%26A...558A..48B} {558, A48}

\bibitem[\protect\citeauthoryear{{Benomar} et~al.,}{{Benomar}
  et~al.}{2018}]{benomar18}
{Benomar} O.,  et~al., 2018, \mn@doi [\apj] {10.3847/1538-4357/aab9b7}, \href
  {http://adsabs.harvard.edu/abs/2018ApJ...857..119B} {857, 119}

\bibitem[\protect\citeauthoryear{{Brown}}{{Brown}}{1991}]{brown91}
{Brown} T.~M.,  1991, \mn@doi [\apj] {10.1086/169900}, \href
  {http://adsabs.harvard.edu/abs/1991ApJ...371..396B} {371, 396}

\bibitem[\protect\citeauthoryear{{Buchhave} \& {Latham}}{{Buchhave} \&
  {Latham}}{2015}]{buchhave15}
{Buchhave} L.~A.,  {Latham} D.~W.,  2015, \mn@doi [\apj]
  {10.1088/0004-637X/808/2/187}, \href
  {http://adsabs.harvard.edu/abs/2015ApJ...808..187B} {808, 187}

\bibitem[\protect\citeauthoryear{{Buchhave} et~al.,}{{Buchhave}
  et~al.}{2012}]{buchhave12}
{Buchhave} L.~A.,  et~al., 2012, \mn@doi [\nat] {10.1038/nature11121}, \href
  {http://adsabs.harvard.edu/abs/2012Natur.486..375B} {486, 375}

\bibitem[\protect\citeauthoryear{{Casagrande} et~al.,}{{Casagrande}
  et~al.}{2014}]{casagrande14}
{Casagrande} L.,  et~al., 2014, \mn@doi [\apj] {10.1088/0004-637X/787/2/110},
  \href {http://adsabs.harvard.edu/abs/2014ApJ...787..110C} {787, 110}

\bibitem[\protect\citeauthoryear{{Chaplin} et~al.,}{{Chaplin}
  et~al.}{2014}]{chaplin14}
{Chaplin} W.~J.,  et~al., 2014, \mn@doi [\apjs] {10.1088/0067-0049/210/1/1},
  \href {http://adsabs.harvard.edu/abs/2014ApJS..210....1C} {210, 1}

\bibitem[\protect\citeauthoryear{{Christensen-Dalsgaard}
  et~al.,}{{Christensen-Dalsgaard} et~al.}{1996}]{cd96}
{Christensen-Dalsgaard} J.,  et~al., 1996, \mn@doi [Science]
  {10.1126/science.272.5266.1286}, \href
  {http://adsabs.harvard.edu/abs/1996Sci...272.1286C} {272, 1286}

\bibitem[\protect\citeauthoryear{{Cox} \& {Giuli}}{{Cox} \&
  {Giuli}}{1968}]{cox68}
{Cox} J.~P.,  {Giuli} R.~T.,  1968, {Principles of stellar structure }

\bibitem[\protect\citeauthoryear{{Cyburt} et~al.,}{{Cyburt}
  et~al.}{2010}]{cyburt10}
{Cyburt} R.~H.,  et~al., 2010, \mn@doi [\apjs] {10.1088/0067-0049/189/1/240},
  \href {http://adsabs.harvard.edu/abs/2010ApJS..189..240C} {189, 240}

\bibitem[\protect\citeauthoryear{{Freytag}, {Ludwig}  \& {Steffen}}{{Freytag}
  et~al.}{1996}]{freytag96}
{Freytag} B.,  {Ludwig} H.-G.,   {Steffen} M.,  1996, \aap, \href
  {http://adsabs.harvard.edu/abs/1996A%26A...313..497F} {313, 497}

\bibitem[\protect\citeauthoryear{{Fr{\"o}hlich}}{{Fr{\"o}hlich}}{2009}]{froehlich09}
{Fr{\"o}hlich} C.,  2009, \mn@doi [\aap] {10.1051/0004-6361/200912318}, \href
  {http://adsabs.harvard.edu/abs/2009A%26A...501L..27F} {501, L27}

\bibitem[\protect\citeauthoryear{{Gough}}{{Gough}}{1987}]{gough87}
{Gough} D.,  1987, \mn@doi [\nat] {10.1038/326257a0}, \href
  {http://adsabs.harvard.edu/abs/1987Natur.326..257G} {326, 257}

\bibitem[\protect\citeauthoryear{{Gough}}{{Gough}}{1990}]{gough90}
{Gough} D.~O.,  1990, in {Osaki} Y.,  {Shibahashi} H.,  eds,  Lecture Notes in
  Physics, Berlin Springer Verlag Vol. 367, Progress of Seismology of the Sun
  and Stars. p.~283, \mn@doi{10.1007/3-540-53091-6}

\bibitem[\protect\citeauthoryear{{Gruberbauer} \& {Guenther}}{{Gruberbauer} \&
  {Guenther}}{2013}]{gruberbauerguenther13}
{Gruberbauer} M.,  {Guenther} D.~B.,  2013, \mn@doi [\mnras]
  {10.1093/mnras/stt477}, \href
  {http://adsabs.harvard.edu/abs/2013MNRAS.432..417G} {432, 417}

\bibitem[\protect\citeauthoryear{{Gruberbauer}, {Guenther}  \&
  {Kallinger}}{{Gruberbauer} et~al.}{2012}]{gruberbauer12}
{Gruberbauer} M.,  {Guenther} D.~B.,   {Kallinger} T.,  2012, \mn@doi [\apj]
  {10.1088/0004-637X/749/2/109}, \href
  {http://adsabs.harvard.edu/abs/2012ApJ...749..109G} {749, 109}

\bibitem[\protect\citeauthoryear{{Gruberbauer}, {Guenther}, {MacLeod}  \&
  {Kallinger}}{{Gruberbauer} et~al.}{2013}]{gruberbauer13}
{Gruberbauer} M.,  {Guenther} D.~B.,  {MacLeod} K.,   {Kallinger} T.,  2013,
  \mn@doi [\mnras] {10.1093/mnras/stt1289}, \href
  {http://adsabs.harvard.edu/abs/2013MNRAS.435..242G} {435, 242}

\bibitem[\protect\citeauthoryear{{Houdek}, {Trampedach}, {Aarslev}  \&
  {Christensen-Dalsgaard}}{{Houdek} et~al.}{2017}]{houdek17}
{Houdek} G.,  {Trampedach} R.,  {Aarslev} M.~J.,   {Christensen-Dalsgaard} J.,
  2017, \mn@doi [\mnras] {10.1093/mnrasl/slw193}, \href
  {http://adsabs.harvard.edu/abs/2017MNRAS.464L.124H} {464, L124}

\bibitem[\protect\citeauthoryear{{Huber} et~al.,}{{Huber}
  et~al.}{2013}]{huber13}
{Huber} D.,  et~al., 2013, \mn@doi [\apj] {10.1088/0004-637X/767/2/127}, \href
  {http://adsabs.harvard.edu/abs/2013ApJ...767..127H} {767, 127}

\bibitem[\protect\citeauthoryear{{Huber} et~al.,}{{Huber}
  et~al.}{2014}]{huber14}
{Huber} D.,  et~al., 2014, \mn@doi [\apjs] {10.1088/0067-0049/211/1/2}, \href
  {http://adsabs.harvard.edu/abs/2014ApJS..211....2H} {211, 2}

\bibitem[\protect\citeauthoryear{{Kjeldsen} \& {Bedding}}{{Kjeldsen} \&
  {Bedding}}{1995}]{kjeldsen95}
{Kjeldsen} H.,  {Bedding} T.~R.,  1995, \aap, \href
  {http://adsabs.harvard.edu/abs/1995A%26A...293...87K} {293, 87}

\bibitem[\protect\citeauthoryear{{Kjeldsen}, {Bedding}  \&
  {Christensen-Dalsgaard}}{{Kjeldsen} et~al.}{2008}]{kjeldsen08}
{Kjeldsen} H.,  {Bedding} T.~R.,   {Christensen-Dalsgaard} J.,  2008, \mn@doi
  [\apjl] {10.1086/591667}, \href
  {http://adsabs.harvard.edu/abs/2008ApJ...683L.175K} {683, L175}

\bibitem[\protect\citeauthoryear{{Kunz}, {Fey}, {Jaeger}, {Mayer}, {Hammer},
  {Staudt}, {Harissopulos}  \& {Paradellis}}{{Kunz} et~al.}{2002}]{kunz02}
{Kunz} R.,  {Fey} M.,  {Jaeger} M.,  {Mayer} A.,  {Hammer} J.~W.,  {Staudt} G.,
   {Harissopulos} S.,   {Paradellis} T.,  2002, \mn@doi [\apj]
  {10.1086/338384}, \href {http://adsabs.harvard.edu/abs/2002ApJ...567..643K}
  {567, 643}

\bibitem[\protect\citeauthoryear{{Lazrek} et~al.,}{{Lazrek}
  et~al.}{1997}]{lazrek97}
{Lazrek} M.,  et~al., 1997, \mn@doi [\solphys] {10.1023/A:1004942929956}, \href
  {http://adsabs.harvard.edu/abs/1997SoPh..175..227L} {175, 227}

\bibitem[\protect\citeauthoryear{{Lebreton} \& {Goupil}}{{Lebreton} \&
  {Goupil}}{2014}]{lebreton14}
{Lebreton} Y.,  {Goupil} M.~J.,  2014, \mn@doi [\aap]
  {10.1051/0004-6361/201423797}, \href
  {http://adsabs.harvard.edu/abs/2014A%26A...569A..21L} {569, A21}

\bibitem[\protect\citeauthoryear{{Li}, {Bedding}, {Huber}, {Ball}, {Stello},
  {Murphy}  \& {Bland-Hawthorn}}{{Li} et~al.}{2018}]{li18}
{Li} T.,  {Bedding} T.~R.,  {Huber} D.,  {Ball} W.~H.,  {Stello} D.,  {Murphy}
  S.~J.,   {Bland-Hawthorn} J.,  2018, \mn@doi [\mnras]
  {10.1093/mnras/stx3079}, \href
  {http://adsabs.harvard.edu/abs/2018MNRAS.475..981L} {475, 981}

\bibitem[\protect\citeauthoryear{{Ludwig}, {Caffau}, {Steffen}, {Freytag},
  {Bonifacio}  \& {Ku{\v c}inskas}}{{Ludwig} et~al.}{2009}]{ludwig09}
{Ludwig} H.-G.,  {Caffau} E.,  {Steffen} M.,  {Freytag} B.,  {Bonifacio} P.,
  {Ku{\v c}inskas} A.,  2009, \memsai, \href
  {http://adsabs.harvard.edu/abs/2009MmSAI..80..711L} {80, 711}

\bibitem[\protect\citeauthoryear{{Lund} et~al.,}{{Lund} et~al.}{2017}]{lund17}
{Lund} M.~N.,  et~al., 2017, \mn@doi [\apj] {10.3847/1538-4357/835/2/172},
  \href {http://adsabs.harvard.edu/abs/2017ApJ...835..172L} {835, 172}

\bibitem[\protect\citeauthoryear{{Magic}, {Weiss}  \& {Asplund}}{{Magic}
  et~al.}{2015}]{magic15}
{Magic} Z.,  {Weiss} A.,   {Asplund} M.,  2015, \mn@doi [\aap]
  {10.1051/0004-6361/201423760}, \href
  {http://adsabs.harvard.edu/abs/2015A%26A...573A..89M} {573, A89}

\bibitem[\protect\citeauthoryear{{Molenda-{\.Z}akowicz}
  et~al.,}{{Molenda-{\.Z}akowicz} et~al.}{2013}]{molenda13}
{Molenda-{\.Z}akowicz} J.,  et~al., 2013, \mn@doi [\mnras]
  {10.1093/mnras/stt1095}, \href
  {http://adsabs.harvard.edu/abs/2013MNRAS.434.1422M} {434, 1422}

\bibitem[\protect\citeauthoryear{{Nsamba}, {Campante}, {Monteiro}, {Cunha},
  {Rendle}, {Reese}  \& {Verma}}{{Nsamba} et~al.}{2018}]{nsamba18}
{Nsamba} B.,  {Campante} T.~L.,  {Monteiro} M.~J.~P.~F.~G.,  {Cunha} M.~S.,
  {Rendle} B.~M.,  {Reese} D.~R.,   {Verma} K.,  2018, \mn@doi [\mnras]
  {10.1093/mnras/sty948}, \href
  {http://adsabs.harvard.edu/abs/2018MNRAS.477.5052N} {477, 5052}

\bibitem[\protect\citeauthoryear{{Paxton}, {Bildsten}, {Dotter}, {Herwig},
  {Lesaffre}  \& {Timmes}}{{Paxton} et~al.}{2011}]{paxton11}
{Paxton} B.,  {Bildsten} L.,  {Dotter} A.,  {Herwig} F.,  {Lesaffre} P.,
  {Timmes} F.,  2011, \mn@doi [\apjs] {10.1088/0067-0049/192/1/3}, \href
  {http://adsabs.harvard.edu/abs/2011ApJS..192....3P} {192, 3}

\bibitem[\protect\citeauthoryear{{Paxton} et~al.,}{{Paxton}
  et~al.}{2013}]{paxton13}
{Paxton} B.,  et~al., 2013, \mn@doi [\apjs] {10.1088/0067-0049/208/1/4}, \href
  {http://adsabs.harvard.edu/abs/2013ApJS..208....4P} {208, 4}

\bibitem[\protect\citeauthoryear{{Paxton} et~al.,}{{Paxton}
  et~al.}{2015}]{paxton15}
{Paxton} B.,  et~al., 2015, \mn@doi [\apjs] {10.1088/0067-0049/220/1/15}, \href
  {http://adsabs.harvard.edu/abs/2015ApJS..220...15P} {220, 15}

\bibitem[\protect\citeauthoryear{{Pinsonneault}, {An}, {Molenda-{\.Z}akowicz},
  {Chaplin}, {Metcalfe}  \& {Bruntt}}{{Pinsonneault}
  et~al.}{2012}]{pinsonneault12}
{Pinsonneault} M.~H.,  {An} D.,  {Molenda-{\.Z}akowicz} J.,  {Chaplin} W.~J.,
  {Metcalfe} T.~S.,   {Bruntt} H.,  2012, \mn@doi [\apjs]
  {10.1088/0067-0049/199/2/30}, \href
  {http://adsabs.harvard.edu/abs/2012ApJS..199...30P} {199, 30}

\bibitem[\protect\citeauthoryear{{Ram{\'{\i}}rez}, {Mel{\'e}ndez}  \&
  {Asplund}}{{Ram{\'{\i}}rez} et~al.}{2009}]{ramirez09}
{Ram{\'{\i}}rez} I.,  {Mel{\'e}ndez} J.,   {Asplund} M.,  2009, \mn@doi [\aap]
  {10.1051/0004-6361/200913038}, \href
  {http://adsabs.harvard.edu/abs/2009A%26A...508L..17R} {508, L17}

\bibitem[\protect\citeauthoryear{{Roxburgh} \& {Vorontsov}}{{Roxburgh} \&
  {Vorontsov}}{2003}]{roxburgh03}
{Roxburgh} I.~W.,  {Vorontsov} S.~V.,  2003, \mn@doi [\aap]
  {10.1051/0004-6361:20031318}, \href
  {http://adsabs.harvard.edu/abs/2003A%26A...411..215R} {411, 215}

\bibitem[\protect\citeauthoryear{{Roxburgh} \& {Vorontsov}}{{Roxburgh} \&
  {Vorontsov}}{2013}]{roxburgh13}
{Roxburgh} I.~W.,  {Vorontsov} S.~V.,  2013, \mn@doi [\aap]
  {10.1051/0004-6361/201321333}, \href
  {http://adsabs.harvard.edu/abs/2013A%26A...560A...2R} {560, A2}

\bibitem[\protect\citeauthoryear{{Schmitt} \& {Basu}}{{Schmitt} \&
  {Basu}}{2015}]{schmittbasu15}
{Schmitt} J.~R.,  {Basu} S.,  2015, \mn@doi [\apj]
  {10.1088/0004-637X/808/2/123}, \href
  {http://adsabs.harvard.edu/abs/2015ApJ...808..123S} {808, 123}

\bibitem[\protect\citeauthoryear{{Silva Aguirre} et~al.,}{{Silva Aguirre}
  et~al.}{2015}]{silva15}
{Silva Aguirre} V.,  et~al., 2015, \mn@doi [\mnras] {10.1093/mnras/stv1388},
  \href {http://adsabs.harvard.edu/abs/2015MNRAS.452.2127S} {452, 2127}

\bibitem[\protect\citeauthoryear{{Silva Aguirre} et~al.,}{{Silva Aguirre}
  et~al.}{2017}]{silva17}
{Silva Aguirre} V.,  et~al., 2017, \mn@doi [\apj]
  {10.3847/1538-4357/835/2/173}, \href
  {http://adsabs.harvard.edu/abs/2017ApJ...835..173S} {835, 173}

\bibitem[\protect\citeauthoryear{{Sonoi}, {Samadi}, {Belkacem}, {Ludwig},
  {Caffau}  \& {Mosser}}{{Sonoi} et~al.}{2015}]{sonoi15}
{Sonoi} T.,  {Samadi} R.,  {Belkacem} K.,  {Ludwig} H.-G.,  {Caffau} E.,
  {Mosser} B.,  2015, \mn@doi [\aap] {10.1051/0004-6361/201526838}, \href
  {http://adsabs.harvard.edu/abs/2015A%26A...583A.112S} {583, A112}

\bibitem[\protect\citeauthoryear{{Tassoul}}{{Tassoul}}{1980}]{tassoul80}
{Tassoul} M.,  1980, \mn@doi [\apjs] {10.1086/190678}, \href
  {http://adsabs.harvard.edu/abs/1980ApJS...43..469T} {43, 469}

\bibitem[\protect\citeauthoryear{{Townsend} \& {Teitler}}{{Townsend} \&
  {Teitler}}{2013}]{townsend13}
{Townsend} R.~H.~D.,  {Teitler} S.~A.,  2013, \mn@doi [\mnras]
  {10.1093/mnras/stt1533}, \href
  {http://adsabs.harvard.edu/abs/2013MNRAS.435.3406T} {435, 3406}

\bibitem[\protect\citeauthoryear{{Trampedach}, {Asplund}, {Collet}, {Nordlund}
  \& {Stein}}{{Trampedach} et~al.}{2013}]{trampedach13}
{Trampedach} R.,  {Asplund} M.,  {Collet} R.,  {Nordlund} {\AA}.,   {Stein}
  R.~F.,  2013, \mn@doi [\apj] {10.1088/0004-637X/769/1/18}, \href
  {http://adsabs.harvard.edu/abs/2013ApJ...769...18T} {769, 18}

\bibitem[\protect\citeauthoryear{{Trampedach}, {Stein},
  {Christensen-Dalsgaard}, {Nordlund}  \& {Asplund}}{{Trampedach}
  et~al.}{2014}]{trampedach14}
{Trampedach} R.,  {Stein} R.~F.,  {Christensen-Dalsgaard} J.,  {Nordlund}
  {\AA}.,   {Asplund} M.,  2014, \mn@doi [\mnras] {10.1093/mnras/stu2084},
  \href {http://adsabs.harvard.edu/abs/2014MNRAS.445.4366T} {445, 4366}

\bibitem[\protect\citeauthoryear{{Trampedach}, {Aarslev}, {Houdek}, {Collet},
  {Christensen-Dalsgaard}, {Stein}  \& {Asplund}}{{Trampedach}
  et~al.}{2017}]{trampedach17}
{Trampedach} R.,  {Aarslev} M.~J.,  {Houdek} G.,  {Collet} R.,
  {Christensen-Dalsgaard} J.,  {Stein} R.~F.,   {Asplund} M.,  2017, \mn@doi
  [\mnras] {10.1093/mnrasl/slw230}, \href
  {https://ui.adsabs.harvard.edu/#abs/2017MNRAS.466L..43T} {466, L43}

\bibitem[\protect\citeauthoryear{{Ulrich}}{{Ulrich}}{1986}]{ulrich86}
{Ulrich} R.~K.,  1986, \mn@doi [\apjl] {10.1086/184700}, \href
  {http://adsabs.harvard.edu/abs/1986ApJ...306L..37U} {306, L37}

\bibitem[\protect\citeauthoryear{{Verma}, {Antia}, {Basu}  \&
  {Mazumdar}}{{Verma} et~al.}{2014}]{verma14b}
{Verma} K.,  {Antia} H.~M.,  {Basu} S.,   {Mazumdar} A.,  2014, \mn@doi [\apj]
  {10.1088/0004-637X/794/2/114}, \href
  {http://adsabs.harvard.edu/abs/2014ApJ...794..114V} {794, 114}

\bibitem[\protect\citeauthoryear{{Verma}, {Raodeo}, {Antia}, {Mazumdar},
  {Basu}, {Lund}  \& {Silva Aguirre}}{{Verma} et~al.}{2017}]{verma17}
{Verma} K.,  {Raodeo} K.,  {Antia} H.~M.,  {Mazumdar} A.,  {Basu} S.,  {Lund}
  M.~N.,   {Silva Aguirre} V.,  2017, \mn@doi [\apj]
  {10.3847/1538-4357/aa5da7}, \href
  {http://adsabs.harvard.edu/abs/2017ApJ...837...47V} {837, 47}

\bibitem[\protect\citeauthoryear{{White} et~al.,}{{White}
  et~al.}{2013}]{white13}
{White} T.~R.,  et~al., 2013, \mn@doi [\mnras] {10.1093/mnras/stt802}, \href
  {http://adsabs.harvard.edu/abs/2013MNRAS.433.1262W} {433, 1262}

\makeatother
\end{thebibliography}

\appendix

\section*{\\Appendix: Tables}

\begin{table*}
\centering
\caption{Stellar and surface correction parameters for LEGACY stars using the power-law correction method.}
\label{tab:table_k}
\begin{tabular}{lrrrrrr}
\hline
\multicolumn{1}{c}{KIC} & \multicolumn{1}{c}{Mass}  & \multicolumn{1}{c}{Initial Z}  & \multicolumn{1}{c}{Radius} &  \multicolumn{1}{c}{Age} & \multicolumn{1}{c}{\change{$\nu_{\rm corr}(\numax) / \nu_{\rm max}$}} & \multicolumn{1}{c}{$r$} \\
& \multicolumn{1}{c}{($M_{\odot}$)} & \multicolumn{1}{c}{($10^{-2}$)} & \multicolumn{1}{c}{($R_{\odot}$)} & \multicolumn{1}{c}{(Gyr)} & \multicolumn{1}{c}{($10^{-3}$)} & \\
\hline
1435467 & 1.324 $\pm$ 0.021 & 1.616 $\pm$ 0.251 & 1.685 $\pm$ 0.017 & 2.957 $\pm$ 0.190 & 2.468 $\pm$ 0.624 & 0.9960 $\pm$ 0.0034 \\
2837475 & 1.392 $\pm$ 0.041 & 1.478 $\pm$ 0.319 & 1.614 $\pm$ 0.018 & 1.803 $\pm$ 0.243 & 1.906 $\pm$ 0.613 & 0.9979 $\pm$ 0.0037 \\
3427720 & 1.098 $\pm$ 0.020 & 1.323 $\pm$ 0.207 & 1.109 $\pm$ 0.009 & 2.751 $\pm$ 0.368 & 4.597 $\pm$ 0.335 & 0.9973 $\pm$ 0.0034 \\
3456181 & 1.570 $\pm$ 0.016 & 1.965 $\pm$ 0.224 & 2.167 $\pm$ 0.019 & 1.864 $\pm$ 0.072 & 1.759 $\pm$ 0.736 & 0.9964 $\pm$ 0.0070 \\
3632418 & 1.441 $\pm$ 0.006 & 2.172 $\pm$ 0.157 & 1.905 $\pm$ 0.008 & 2.531 $\pm$ 0.050 & 2.754 $\pm$ 0.222 & 0.9864 $\pm$ 0.0045 \\
3656476 & 1.043 $\pm$ 0.014 & 2.190 $\pm$ 0.249 & 1.292 $\pm$ 0.009 & 8.770 $\pm$ 0.483 & 5.135 $\pm$ 0.334 & 0.9924 $\pm$ 0.0048 \\
3735871 & 1.111 $\pm$ 0.029 & 1.308 $\pm$ 0.244 & 1.097 $\pm$ 0.010 & 2.097 $\pm$ 0.679 & 4.188 $\pm$ 0.439 & 0.9989 $\pm$ 0.0016 \\
4914923 & 1.090 $\pm$ 0.019 & 1.868 $\pm$ 0.229 & 1.357 $\pm$ 0.012 & 6.875 $\pm$ 0.377 & 4.998 $\pm$ 0.308 & 0.9923 $\pm$ 0.0044 \\
5184732 & 1.185 $\pm$ 0.011 & 2.912 $\pm$ 0.341 & 1.322 $\pm$ 0.007 & 4.318 $\pm$ 0.251 & 4.735 $\pm$ 0.479 & 0.9900 $\pm$ 0.0047 \\
5773345 & 1.502 $\pm$ 0.020 & 2.373 $\pm$ 0.342 & 2.018 $\pm$ 0.017 & 2.243 $\pm$ 0.104 & 3.388 $\pm$ 0.375 & 0.9974 $\pm$ 0.0058 \\
5950854 & 0.970 $\pm$ 0.022 & 1.032 $\pm$ 0.160 & 1.232 $\pm$ 0.011 & 9.128 $\pm$ 0.799 & 4.549 $\pm$ 0.398 & 0.9959 $\pm$ 0.0026 \\
6106415 & 1.078 $\pm$ 0.017 & 1.434 $\pm$ 0.234 & 1.213 $\pm$ 0.010 & 5.220 $\pm$ 0.365 & 4.079 $\pm$ 0.432 & 0.9942 $\pm$ 0.0038 \\
6116048 & 1.026 $\pm$ 0.016 & 1.087 $\pm$ 0.141 & 1.218 $\pm$ 0.010 & 6.658 $\pm$ 0.434 & 3.989 $\pm$ 0.354 & 0.9914 $\pm$ 0.0042 \\
6225718 & 1.150 $\pm$ 0.021 & 1.452 $\pm$ 0.266 & 1.226 $\pm$ 0.010 & 3.089 $\pm$ 0.318 & 3.260 $\pm$ 0.484 & 0.9943 $\pm$ 0.0030 \\
6508366 & 1.590 $\pm$ 0.013 & 2.362 $\pm$ 0.297 & 2.195 $\pm$ 0.016 & 1.868 $\pm$ 0.030 & 1.499 $\pm$ 0.387 & 0.9908 $\pm$ 0.0052 \\
6603624 & 1.007 $\pm$ 0.011 & 2.133 $\pm$ 0.223 & 1.136 $\pm$ 0.009 & 8.611 $\pm$ 0.395 & 4.650 $\pm$ 0.303 & 0.9872 $\pm$ 0.0064 \\
6679371 & 1.616 $\pm$ 0.010 & 2.015 $\pm$ 0.221 & 2.243 $\pm$ 0.019 & 1.693 $\pm$ 0.037 & 0.385 $\pm$ 0.686 & 0.9984 $\pm$ 0.0044 \\
6933899 & 1.167 $\pm$ 0.012 & 2.120 $\pm$ 0.060 & 1.576 $\pm$ 0.014 & 5.949 $\pm$ 0.242 & 4.993 $\pm$ 0.256 & 0.9752 $\pm$ 0.0094 \\
7103006 & 1.474 $\pm$ 0.021 & 1.923 $\pm$ 0.229 & 1.956 $\pm$ 0.016 & 2.212 $\pm$ 0.136 & 1.172 $\pm$ 0.418 & 0.9986 $\pm$ 0.0043 \\
7106245 & 0.941 $\pm$ 0.025 & 0.479 $\pm$ 0.101 & 1.099 $\pm$ 0.012 & 6.828 $\pm$ 0.748 & 6.753 $\pm$ 0.365 & 0.9949 $\pm$ 0.0031 \\
7206837 & 1.324 $\pm$ 0.035 & 1.871 $\pm$ 0.386 & 1.564 $\pm$ 0.015 & 2.605 $\pm$ 0.340 & 2.204 $\pm$ 0.674 & 0.9991 $\pm$ 0.0048 \\
7296438 & 1.100 $\pm$ 0.021 & 2.120 $\pm$ 0.307 & 1.365 $\pm$ 0.012 & 6.905 $\pm$ 0.548 & 5.307 $\pm$ 0.423 & 0.9974 $\pm$ 0.0038 \\
7510397 & 1.440 $\pm$ 0.005 & 2.450 $\pm$ 0.261 & 1.891 $\pm$ 0.008 & 2.594 $\pm$ 0.067 & 3.035 $\pm$ 0.522 & 0.9989 $\pm$ 0.0068 \\
7680114 & 1.073 $\pm$ 0.019 & 1.778 $\pm$ 0.235 & 1.390 $\pm$ 0.011 & 7.549 $\pm$ 0.525 & 5.163 $\pm$ 0.452 & 0.9953 $\pm$ 0.0042 \\
7771282 & 1.227 $\pm$ 0.059 & 1.408 $\pm$ 0.291 & 1.619 $\pm$ 0.025 & 4.060 $\pm$ 0.756 & 3.148 $\pm$ 0.620 & 0.9995 $\pm$ 0.0032 \\
7871531 & 0.858 $\pm$ 0.014 & 0.979 $\pm$ 0.147 & 0.876 $\pm$ 0.006 & 9.181 $\pm$ 0.925 & 2.876 $\pm$ 0.233 & 0.9963 $\pm$ 0.0026 \\
7940546 & 1.479 $\pm$ 0.006 & 2.117 $\pm$ 0.053 & 1.980 $\pm$ 0.010 & 2.314 $\pm$ 0.038 & 1.844 $\pm$ 0.111 & 0.9989 $\pm$ 0.0058 \\
7970740 & 0.840 $\pm$ 0.000 & 1.086 $\pm$ 0.051 & 0.811 $\pm$ 0.001 & 7.245 $\pm$ 0.188 & 1.316 $\pm$ 0.049 & 1.0243 $\pm$ 0.0027 \\
8006161 & 0.960 $\pm$ 0.010 & 2.687 $\pm$ 0.414 & 0.919 $\pm$ 0.007 & 5.308 $\pm$ 0.377 & 3.021 $\pm$ 0.325 & 0.9979 $\pm$ 0.0061 \\
8150065 & 1.172 $\pm$ 0.045 & 1.324 $\pm$ 0.353 & 1.382 $\pm$ 0.019 & 3.893 $\pm$ 0.652 & 5.582 $\pm$ 0.859 & 0.9985 $\pm$ 0.0022 \\
8179536 & 1.198 $\pm$ 0.039 & 1.347 $\pm$ 0.275 & 1.334 $\pm$ 0.015 & 2.954 $\pm$ 0.635 & 3.509 $\pm$ 0.583 & 0.9985 $\pm$ 0.0015 \\
8228742 & 1.441 $\pm$ 0.004 & 2.581 $\pm$ 0.179 & 1.888 $\pm$ 0.006 & 2.618 $\pm$ 0.054 & 3.842 $\pm$ 0.365 & 0.9951 $\pm$ 0.0049 \\
8379927 & 1.130 $\pm$ 0.020 & 1.472 $\pm$ 0.331 & 1.118 $\pm$ 0.010 & 2.019 $\pm$ 0.308 & 3.377 $\pm$ 0.521 & 0.9950 $\pm$ 0.0036 \\
8394589 & 1.032 $\pm$ 0.029 & 0.836 $\pm$ 0.154 & 1.156 $\pm$ 0.013 & 4.969 $\pm$ 0.666 & 4.149 $\pm$ 0.461 & 0.9973 $\pm$ 0.0021 \\
8424992 & 0.922 $\pm$ 0.020 & 1.158 $\pm$ 0.201 & 1.043 $\pm$ 0.010 & 9.858 $\pm$ 1.051 & 4.492 $\pm$ 0.396 & 0.9964 $\pm$ 0.0028 \\
8694723 & 1.135 $\pm$ 0.016 & 1.042 $\pm$ 0.129 & 1.529 $\pm$ 0.016 & 5.094 $\pm$ 0.200 & 2.484 $\pm$ 0.458 & 0.9839 $\pm$ 0.0086 \\
8760414 & 0.840 $\pm$ 0.001 & 0.435 $\pm$ 0.022 & 1.034 $\pm$ 0.001 & 11.653 $\pm$ 0.145 & 3.425 $\pm$ 0.097 & 0.9981 $\pm$ 0.0006 \\
8938364 & 0.969 $\pm$ 0.014 & 1.324 $\pm$ 0.215 & 1.338 $\pm$ 0.010 & 10.868 $\pm$ 0.620 & 5.522 $\pm$ 0.493 & 0.9957 $\pm$ 0.0045 \\
9025370 & 0.992 $\pm$ 0.022 & 1.175 $\pm$ 0.359 & 1.004 $\pm$ 0.010 & 4.669 $\pm$ 0.531 & 4.297 $\pm$ 0.591 & 0.9991 $\pm$ 0.0034 \\
9098294 & 0.974 $\pm$ 0.018 & 1.148 $\pm$ 0.158 & 1.138 $\pm$ 0.010 & 8.273 $\pm$ 0.714 & 4.568 $\pm$ 0.305 & 0.9949 $\pm$ 0.0042 \\
9139151 & 1.135 $\pm$ 0.025 & 1.603 $\pm$ 0.277 & 1.141 $\pm$ 0.010 & 2.424 $\pm$ 0.517 & 4.404 $\pm$ 0.455 & 0.9990 $\pm$ 0.0021 \\
9139163 & 1.362 $\pm$ 0.029 & 2.368 $\pm$ 0.492 & 1.548 $\pm$ 0.014 & 2.151 $\pm$ 0.216 & 1.357 $\pm$ 0.666 & 0.9936 $\pm$ 0.0065 \\
9206432 & 1.382 $\pm$ 0.037 & 1.951 $\pm$ 0.406 & 1.502 $\pm$ 0.015 & 1.509 $\pm$ 0.276 & 2.170 $\pm$ 0.572 & 0.9998 $\pm$ 0.0021 \\
9353712 & 1.587 $\pm$ 0.011 & 2.234 $\pm$ 0.252 & 2.184 $\pm$ 0.016 & 1.843 $\pm$ 0.044 & 5.774 $\pm$ 0.622 & 0.9939 $\pm$ 0.0074 \\
9410862 & 0.967 $\pm$ 0.028 & 0.758 $\pm$ 0.144 & 1.147 $\pm$ 0.013 & 7.358 $\pm$ 0.949 & 4.089 $\pm$ 0.516 & 0.9978 $\pm$ 0.0021 \\
9414417 & 1.458 $\pm$ 0.015 & 2.064 $\pm$ 0.147 & 1.936 $\pm$ 0.016 & 2.387 $\pm$ 0.118 & 1.988 $\pm$ 0.375 & 0.9944 $\pm$ 0.0076 \\
9812850 & 1.401 $\pm$ 0.027 & 1.549 $\pm$ 0.226 & 1.813 $\pm$ 0.022 & 2.383 $\pm$ 0.124 & 2.045 $\pm$ 0.446 & 0.9999 $\pm$ 0.0069 \\
9955598 & 0.901 $\pm$ 0.013 & 1.485 $\pm$ 0.299 & 0.884 $\pm$ 0.007 & 6.956 $\pm$ 0.681 & 3.227 $\pm$ 0.358 & 0.9991 $\pm$ 0.0047 \\
9965715 & 1.090 $\pm$ 0.046 & 0.794 $\pm$ 0.277 & 1.271 $\pm$ 0.022 & 4.228 $\pm$ 0.677 & 3.357 $\pm$ 0.839 & 0.9981 $\pm$ 0.0019 \\
10068307 & 1.549 $\pm$ 0.024 & 2.588 $\pm$ 0.268 & 2.104 $\pm$ 0.034 & 2.063 $\pm$ 0.084 & 4.420 $\pm$ 0.252 & 0.9885 $\pm$ 0.0163 \\
10079226 & 1.110 $\pm$ 0.044 & 1.865 $\pm$ 0.342 & 1.141 $\pm$ 0.015 & 3.594 $\pm$ 1.283 & 3.952 $\pm$ 0.423 & 0.9989 $\pm$ 0.0014 \\
10162436 & 1.496 $\pm$ 0.011 & 2.131 $\pm$ 0.079 & 2.024 $\pm$ 0.015 & 2.255 $\pm$ 0.056 & 4.190 $\pm$ 0.161 & 0.9796 $\pm$ 0.0080 \\
10454113 & 1.185 $\pm$ 0.027 & 1.533 $\pm$ 0.231 & 1.242 $\pm$ 0.012 & 2.446 $\pm$ 0.431 & 3.057 $\pm$ 0.291 & 0.9938 $\pm$ 0.0053 \\
10516096 & 1.065 $\pm$ 0.019 & 1.281 $\pm$ 0.209 & 1.393 $\pm$ 0.011 & 6.930 $\pm$ 0.446 & 4.907 $\pm$ 0.460 & 0.9947 $\pm$ 0.0036 \\
10644253 & 1.146 $\pm$ 0.022 & 1.620 $\pm$ 0.284 & 1.109 $\pm$ 0.009 & 1.480 $\pm$ 0.409 & 4.173 $\pm$ 0.471 & 0.9990 $\pm$ 0.0030 \\
10730618 & 1.382 $\pm$ 0.045 & 1.943 $\pm$ 0.629 & 1.779 $\pm$ 0.030 & 2.656 $\pm$ 0.335 & 2.427 $\pm$ 1.134 & 0.9972 $\pm$ 0.0054 \\
10963065 & 1.058 $\pm$ 0.021 & 1.022 $\pm$ 0.173 & 1.214 $\pm$ 0.011 & 5.205 $\pm$ 0.452 & 4.057 $\pm$ 0.497 & 0.9971 $\pm$ 0.0035 \\
11081729 & 1.280 $\pm$ 0.046 & 1.720 $\pm$ 0.362 & 1.418 $\pm$ 0.016 & 2.347 $\pm$ 0.566 & 1.237 $\pm$ 1.094 & 0.9994 $\pm$ 0.0016 \\
11253226 & 1.369 $\pm$ 0.034 & 1.317 $\pm$ 0.255 & 1.590 $\pm$ 0.017 & 1.876 $\pm$ 0.190 & 1.035 $\pm$ 0.384 & 0.9991 $\pm$ 0.0038 \\
11772920 & 0.853 $\pm$ 0.013 & 1.282 $\pm$ 0.415 & 0.853 $\pm$ 0.007 & 9.219 $\pm$ 0.957 & 2.914 $\pm$ 0.501 & 1.0000 $\pm$ 0.0041 \\
12009504 & 1.151 $\pm$ 0.028 & 1.385 $\pm$ 0.277 & 1.385 $\pm$ 0.015 & 4.474 $\pm$ 0.416 & 4.294 $\pm$ 0.609 & 0.9941 $\pm$ 0.0031 \\
12069127 & 1.628 $\pm$ 0.010 & 2.138 $\pm$ 0.146 & 2.297 $\pm$ 0.009 & 1.732 $\pm$ 0.050 & 4.489 $\pm$ 0.418 & 0.9962 $\pm$ 0.0043 \\
12069424 & 1.030 $\pm$ 0.010 & 1.690 $\pm$ 0.013 & 1.195 $\pm$ 0.007 & 7.498 $\pm$ 0.407 & 4.490 $\pm$ 0.162 & 0.9865 $\pm$ 0.0042 \\
12069449 & 0.990 $\pm$ 0.018 & 1.608 $\pm$ 0.145 & 1.086 $\pm$ 0.012 & 7.623 $\pm$ 0.487 & 4.233 $\pm$ 0.087 & 0.9881 $\pm$ 0.0075 \\
12258514 & 1.176 $\pm$ 0.013 & 1.902 $\pm$ 0.224 & 1.556 $\pm$ 0.011 & 5.376 $\pm$ 0.236 & 5.277 $\pm$ 0.479 & 0.9899 $\pm$ 0.0061 \\
12317678 & 1.405 $\pm$ 0.013 & 1.111 $\pm$ 0.086 & 1.832 $\pm$ 0.008 & 2.156 $\pm$ 0.055 & 1.750 $\pm$ 0.154 & 1.0037 $\pm$ 0.0023 \\
Sun & 0.989 $\pm$ 0.010 & 1.350 $\pm$ 0.001 & 0.990 $\pm$ 0.007 & 4.824 $\pm$ 0.364 & 3.836 $\pm$ 0.103 & 0.9954 $\pm$ 0.0060 \\
\hline
\end{tabular}
\end{table*}

\begin{table*}
\centering
\caption{Stellar and surface correction parameters for LEGACY stars using the cubic correction method.}
\label{tab:table_c}
\begin{tabular}{lrrrrrr}
\hline
\multicolumn{1}{c}{KIC} & \multicolumn{1}{c}{Mass}  & \multicolumn{1}{c}{Initial Z}  & \multicolumn{1}{c}{Radius} &  \multicolumn{1}{c}{Age} & \multicolumn{1}{c}{\change{$\nu_{\rm corr}(\numax) / \nu_{\rm max}$}} & \multicolumn{1}{c}{$r$} \\
& \multicolumn{1}{c}{($M_{\odot}$)} & \multicolumn{1}{c}{($10^{-2}$)} & \multicolumn{1}{c}{($R_{\odot}$)} & \multicolumn{1}{c}{(Gyr)} & \multicolumn{1}{c}{($10^{-3}$)} & \\
\hline
1435467 & 1.320 $\pm$ 0.022 & 1.542 $\pm$ 0.238 & 1.681 $\pm$ 0.016 & 2.945 $\pm$ 0.198 & 3.006 $\pm$ 0.881 & 0.9960 $\pm$ 0.0037 \\
2837475 & 1.395 $\pm$ 0.038 & 1.437 $\pm$ 0.303 & 1.611 $\pm$ 0.018 & 1.744 $\pm$ 0.217 & 2.359 $\pm$ 0.730 & 0.9973 $\pm$ 0.0048 \\
3427720 & 1.098 $\pm$ 0.020 & 1.311 $\pm$ 0.203 & 1.110 $\pm$ 0.009 & 2.773 $\pm$ 0.360 & 2.953 $\pm$ 0.258 & 0.9976 $\pm$ 0.0036 \\
3456181 & 1.566 $\pm$ 0.016 & 1.881 $\pm$ 0.222 & 2.162 $\pm$ 0.019 & 1.851 $\pm$ 0.071 & 2.474 $\pm$ 0.562 & 0.9970 $\pm$ 0.0072 \\
3632418 & 1.442 $\pm$ 0.006 & 2.130 $\pm$ 0.075 & 1.910 $\pm$ 0.007 & 2.528 $\pm$ 0.045 & 3.005 $\pm$ 0.133 & 0.9906 $\pm$ 0.0035 \\
3656476 & 1.047 $\pm$ 0.014 & 2.205 $\pm$ 0.248 & 1.296 $\pm$ 0.010 & 8.666 $\pm$ 0.476 & 3.014 $\pm$ 0.254 & 0.9926 $\pm$ 0.0056 \\
3735871 & 1.109 $\pm$ 0.029 & 1.286 $\pm$ 0.240 & 1.097 $\pm$ 0.010 & 2.153 $\pm$ 0.667 & 2.879 $\pm$ 0.364 & 0.9994 $\pm$ 0.0015 \\
4914923 & 1.091 $\pm$ 0.019 & 1.873 $\pm$ 0.228 & 1.358 $\pm$ 0.012 & 6.848 $\pm$ 0.368 & 3.116 $\pm$ 0.233 & 0.9916 $\pm$ 0.0050 \\
5184732 & 1.185 $\pm$ 0.012 & 2.794 $\pm$ 0.321 & 1.325 $\pm$ 0.011 & 4.305 $\pm$ 0.224 & 3.133 $\pm$ 0.417 & 0.9920 $\pm$ 0.0090 \\
5773345 & 1.501 $\pm$ 0.020 & 2.294 $\pm$ 0.292 & 2.019 $\pm$ 0.015 & 2.236 $\pm$ 0.112 & 3.713 $\pm$ 0.453 & 0.9994 $\pm$ 0.0052 \\
5950854 & 0.972 $\pm$ 0.022 & 1.035 $\pm$ 0.160 & 1.235 $\pm$ 0.012 & 9.081 $\pm$ 0.779 & 2.824 $\pm$ 0.306 & 0.9967 $\pm$ 0.0032 \\
6106415 & 1.074 $\pm$ 0.016 & 1.299 $\pm$ 0.211 & 1.214 $\pm$ 0.009 & 5.140 $\pm$ 0.343 & 3.079 $\pm$ 0.364 & 0.9970 $\pm$ 0.0040 \\
6116048 & 1.024 $\pm$ 0.017 & 1.026 $\pm$ 0.138 & 1.219 $\pm$ 0.010 & 6.556 $\pm$ 0.404 & 2.980 $\pm$ 0.339 & 0.9933 $\pm$ 0.0043 \\
6225718 & 1.141 $\pm$ 0.020 & 1.236 $\pm$ 0.228 & 1.223 $\pm$ 0.010 & 3.059 $\pm$ 0.296 & 3.022 $\pm$ 0.462 & 0.9969 $\pm$ 0.0029 \\
6508366 & 1.585 $\pm$ 0.012 & 2.230 $\pm$ 0.223 & 2.189 $\pm$ 0.015 & 1.866 $\pm$ 0.031 & 2.496 $\pm$ 0.332 & 0.9907 $\pm$ 0.0053 \\
6603624 & 1.010 $\pm$ 0.011 & 2.139 $\pm$ 0.232 & 1.139 $\pm$ 0.009 & 8.512 $\pm$ 0.388 & 2.565 $\pm$ 0.237 & 0.9877 $\pm$ 0.0071 \\
6679371 & 1.611 $\pm$ 0.011 & 1.855 $\pm$ 0.227 & 2.231 $\pm$ 0.020 & 1.660 $\pm$ 0.051 & 1.439 $\pm$ 1.034 & 0.9983 $\pm$ 0.0056 \\
6933899 & 1.168 $\pm$ 0.011 & 2.120 $\pm$ 0.047 & 1.575 $\pm$ 0.012 & 5.936 $\pm$ 0.234 & 3.189 $\pm$ 0.204 & 0.9727 $\pm$ 0.0078 \\
7103006 & 1.469 $\pm$ 0.022 & 1.803 $\pm$ 0.196 & 1.950 $\pm$ 0.016 & 2.189 $\pm$ 0.145 & 2.813 $\pm$ 0.713 & 0.9991 $\pm$ 0.0045 \\
7106245 & 0.936 $\pm$ 0.025 & 0.471 $\pm$ 0.102 & 1.099 $\pm$ 0.012 & 6.982 $\pm$ 0.734 & 4.623 $\pm$ 0.329 & 0.9945 $\pm$ 0.0032 \\
7206837 & 1.326 $\pm$ 0.035 & 1.842 $\pm$ 0.376 & 1.565 $\pm$ 0.015 & 2.570 $\pm$ 0.340 & 2.272 $\pm$ 0.753 & 1.0002 $\pm$ 0.0058 \\
7296438 & 1.102 $\pm$ 0.021 & 2.136 $\pm$ 0.307 & 1.367 $\pm$ 0.011 & 6.866 $\pm$ 0.540 & 3.336 $\pm$ 0.334 & 0.9968 $\pm$ 0.0038 \\
7510397 & 1.437 $\pm$ 0.010 & 2.178 $\pm$ 0.231 & 1.892 $\pm$ 0.014 & 2.537 $\pm$ 0.053 & 3.185 $\pm$ 0.468 & 1.0027 $\pm$ 0.0070 \\
7680114 & 1.074 $\pm$ 0.019 & 1.735 $\pm$ 0.213 & 1.392 $\pm$ 0.012 & 7.461 $\pm$ 0.506 & 3.390 $\pm$ 0.321 & 0.9959 $\pm$ 0.0045 \\
7771282 & 1.230 $\pm$ 0.059 & 1.409 $\pm$ 0.288 & 1.620 $\pm$ 0.025 & 4.014 $\pm$ 0.761 & 2.950 $\pm$ 0.687 & 0.9995 $\pm$ 0.0033 \\
7871531 & 0.855 $\pm$ 0.013 & 0.929 $\pm$ 0.117 & 0.876 $\pm$ 0.005 & 9.256 $\pm$ 0.909 & 1.724 $\pm$ 0.137 & 0.9965 $\pm$ 0.0035 \\
7940546 & 1.478 $\pm$ 0.008 & 2.081 $\pm$ 0.125 & 1.981 $\pm$ 0.012 & 2.309 $\pm$ 0.042 & 2.545 $\pm$ 0.365 & 1.0008 $\pm$ 0.0054 \\
7970740 & 0.840 $\pm$ 0.000 & 1.085 $\pm$ 0.046 & 0.812 $\pm$ 0.001 & 7.331 $\pm$ 0.167 & 0.816 $\pm$ 0.034 & 1.0257 $\pm$ 0.0025 \\
8006161 & 0.961 $\pm$ 0.010 & 2.617 $\pm$ 0.437 & 0.921 $\pm$ 0.007 & 5.293 $\pm$ 0.386 & 1.653 $\pm$ 0.241 & 0.9994 $\pm$ 0.0061 \\
8150065 & 1.170 $\pm$ 0.045 & 1.312 $\pm$ 0.354 & 1.382 $\pm$ 0.020 & 3.925 $\pm$ 0.645 & 4.384 $\pm$ 0.867 & 0.9980 $\pm$ 0.0021 \\
8179536 & 1.201 $\pm$ 0.040 & 1.337 $\pm$ 0.269 & 1.334 $\pm$ 0.015 & 2.907 $\pm$ 0.650 & 3.067 $\pm$ 0.612 & 0.9985 $\pm$ 0.0016 \\
8228742 & 1.441 $\pm$ 0.004 & 2.424 $\pm$ 0.262 & 1.891 $\pm$ 0.006 & 2.579 $\pm$ 0.074 & 3.374 $\pm$ 0.482 & 0.9982 $\pm$ 0.0053 \\
8379927 & 1.120 $\pm$ 0.021 & 1.217 $\pm$ 0.285 & 1.116 $\pm$ 0.010 & 2.048 $\pm$ 0.293 & 2.953 $\pm$ 0.503 & 0.9984 $\pm$ 0.0033 \\
8394589 & 1.031 $\pm$ 0.028 & 0.821 $\pm$ 0.152 & 1.156 $\pm$ 0.012 & 4.992 $\pm$ 0.639 & 3.131 $\pm$ 0.414 & 0.9977 $\pm$ 0.0016 \\
8424992 & 0.923 $\pm$ 0.020 & 1.160 $\pm$ 0.201 & 1.045 $\pm$ 0.010 & 9.824 $\pm$ 1.030 & 2.422 $\pm$ 0.257 & 0.9966 $\pm$ 0.0028 \\
8694723 & 1.130 $\pm$ 0.017 & 0.950 $\pm$ 0.120 & 1.526 $\pm$ 0.017 & 5.001 $\pm$ 0.188 & 2.742 $\pm$ 0.492 & 0.9853 $\pm$ 0.0083 \\
8760414 & 0.840 $\pm$ 0.001 & 0.433 $\pm$ 0.016 & 1.034 $\pm$ 0.001 & 11.64 $\pm$ 0.111 & 2.353 $\pm$ 0.049 & 0.9971 $\pm$ 0.0006 \\
8938364 & 0.971 $\pm$ 0.014 & 1.308 $\pm$ 0.206 & 1.340 $\pm$ 0.011 & 10.743 $\pm$ 0.586 & 3.435 $\pm$ 0.350 & 0.9952 $\pm$ 0.0051 \\
9025370 & 0.991 $\pm$ 0.022 & 1.138 $\pm$ 0.357 & 1.005 $\pm$ 0.010 & 4.637 $\pm$ 0.517 & 2.573 $\pm$ 0.469 & 0.9995 $\pm$ 0.0034 \\
9098294 & 0.977 $\pm$ 0.018 & 1.145 $\pm$ 0.151 & 1.140 $\pm$ 0.010 & 8.147 $\pm$ 0.687 & 2.891 $\pm$ 0.234 & 0.9954 $\pm$ 0.0047 \\
9139151 & 1.132 $\pm$ 0.024 & 1.559 $\pm$ 0.279 & 1.141 $\pm$ 0.010 & 2.462 $\pm$ 0.495 & 3.032 $\pm$ 0.402 & 0.9997 $\pm$ 0.0021 \\
9139163 & 1.339 $\pm$ 0.029 & 1.743 $\pm$ 0.384 & 1.535 $\pm$ 0.018 & 2.155 $\pm$ 0.183 & 2.961 $\pm$ 0.809 & 0.9981 $\pm$ 0.0074 \\
9206432 & 1.391 $\pm$ 0.038 & 1.947 $\pm$ 0.401 & 1.502 $\pm$ 0.016 & 1.405 $\pm$ 0.268 & 2.690 $\pm$ 0.829 & 0.9990 $\pm$ 0.0020 \\
9353712 & 1.587 $\pm$ 0.011 & 2.213 $\pm$ 0.223 & 2.184 $\pm$ 0.016 & 1.840 $\pm$ 0.043 & 4.293 $\pm$ 0.348 & 0.9926 $\pm$ 0.0074 \\
9410862 & 0.968 $\pm$ 0.028 & 0.759 $\pm$ 0.144 & 1.149 $\pm$ 0.013 & 7.326 $\pm$ 0.916 & 2.815 $\pm$ 0.423 & 0.9983 $\pm$ 0.0027 \\
9414417 & 1.457 $\pm$ 0.015 & 1.950 $\pm$ 0.211 & 1.934 $\pm$ 0.017 & 2.343 $\pm$ 0.118 & 2.722 $\pm$ 0.679 & 0.9964 $\pm$ 0.0076 \\
9812850 & 1.394 $\pm$ 0.026 & 1.491 $\pm$ 0.197 & 1.809 $\pm$ 0.021 & 2.409 $\pm$ 0.130 & 2.878 $\pm$ 0.540 & 1.0005 $\pm$ 0.0068 \\
9955598 & 0.902 $\pm$ 0.013 & 1.485 $\pm$ 0.296 & 0.886 $\pm$ 0.007 & 6.923 $\pm$ 0.667 & 1.742 $\pm$ 0.230 & 0.9991 $\pm$ 0.0050 \\
9965715 & 1.083 $\pm$ 0.046 & 0.732 $\pm$ 0.254 & 1.268 $\pm$ 0.022 & 4.265 $\pm$ 0.678 & 3.186 $\pm$ 0.865 & 0.9984 $\pm$ 0.0016 \\
10068307 & 1.551 $\pm$ 0.023 & 2.571 $\pm$ 0.212 & 2.108 $\pm$ 0.034 & 2.057 $\pm$ 0.081 & 3.740 $\pm$ 0.191 & 0.9903 $\pm$ 0.0164 \\
10079226 & 1.106 $\pm$ 0.044 & 1.822 $\pm$ 0.338 & 1.141 $\pm$ 0.015 & 3.706 $\pm$ 1.258 & 2.820 $\pm$ 0.404 & 0.9993 $\pm$ 0.0015 \\
10162436 & 1.498 $\pm$ 0.008 & 2.125 $\pm$ 0.049 & 2.029 $\pm$ 0.011 & 2.248 $\pm$ 0.043 & 4.250 $\pm$ 0.171 & 0.9826 $\pm$ 0.0060 \\
10454113 & 1.184 $\pm$ 0.028 & 1.514 $\pm$ 0.225 & 1.241 $\pm$ 0.012 & 2.451 $\pm$ 0.441 & 2.450 $\pm$ 0.268 & 0.9933 $\pm$ 0.0054 \\
10516096 & 1.065 $\pm$ 0.018 & 1.258 $\pm$ 0.200 & 1.395 $\pm$ 0.011 & 6.903 $\pm$ 0.412 & 3.448 $\pm$ 0.386 & 0.9956 $\pm$ 0.0036 \\
10644253 & 1.147 $\pm$ 0.022 & 1.616 $\pm$ 0.281 & 1.111 $\pm$ 0.009 & 1.486 $\pm$ 0.397 & 2.735 $\pm$ 0.370 & 0.9992 $\pm$ 0.0032 \\
10730618 & 1.375 $\pm$ 0.042 & 1.752 $\pm$ 0.587 & 1.773 $\pm$ 0.028 & 2.631 $\pm$ 0.308 & 2.745 $\pm$ 1.042 & 0.9970 $\pm$ 0.0053 \\
10963065 & 1.056 $\pm$ 0.022 & 0.994 $\pm$ 0.171 & 1.215 $\pm$ 0.011 & 5.212 $\pm$ 0.445 & 2.924 $\pm$ 0.426 & 0.9980 $\pm$ 0.0028 \\
11081729 & 1.291 $\pm$ 0.047 & 1.692 $\pm$ 0.359 & 1.419 $\pm$ 0.016 & 2.166 $\pm$ 0.586 & 1.700 $\pm$ 1.393 & 0.9996 $\pm$ 0.0019 \\
11253226 & 1.368 $\pm$ 0.035 & 1.260 $\pm$ 0.248 & 1.587 $\pm$ 0.017 & 1.856 $\pm$ 0.206 & 1.820 $\pm$ 0.647 & 1.0003 $\pm$ 0.0037 \\
11772920 & 0.852 $\pm$ 0.013 & 1.223 $\pm$ 0.386 & 0.854 $\pm$ 0.006 & 9.186 $\pm$ 0.935 & 1.631 $\pm$ 0.316 & 1.0006 $\pm$ 0.0038 \\
12009504 & 1.150 $\pm$ 0.027 & 1.284 $\pm$ 0.239 & 1.385 $\pm$ 0.014 & 4.381 $\pm$ 0.380 & 3.740 $\pm$ 0.539 & 0.9963 $\pm$ 0.0030 \\
12069127 & 1.628 $\pm$ 0.010 & 2.132 $\pm$ 0.120 & 2.297 $\pm$ 0.009 & 1.731 $\pm$ 0.050 & 3.412 $\pm$ 0.212 & 0.9953 $\pm$ 0.0042 \\
12069424 & 1.035 $\pm$ 0.009 & 1.690 $\pm$ 0.013 & 1.200 $\pm$ 0.007 & 7.336 $\pm$ 0.333 & 2.836 $\pm$ 0.128 & 0.9888 $\pm$ 0.0047 \\
12069449 & 0.989 $\pm$ 0.019 & 1.597 $\pm$ 0.152 & 1.088 $\pm$ 0.012 & 7.694 $\pm$ 0.501 & 2.510 $\pm$ 0.056 & 0.9897 $\pm$ 0.0074 \\
12258514 & 1.177 $\pm$ 0.012 & 1.811 $\pm$ 0.196 & 1.558 $\pm$ 0.011 & 5.255 $\pm$ 0.220 & 4.111 $\pm$ 0.439 & 0.9915 $\pm$ 0.0071 \\
12317678 & 1.403 $\pm$ 0.010 & 1.096 $\pm$ 0.063 & 1.827 $\pm$ 0.007 & 2.149 $\pm$ 0.046 & 2.093 $\pm$ 0.220 & 1.0007 $\pm$ 0.0027 \\
Sun & 0.993 $\pm$ 0.010 & 1.350 $\pm$ 0.001 & 0.996 $\pm$ 0.007 & 4.743 $\pm$ 0.348 & 2.354 $\pm$ 0.098 & 0.9999 $\pm$ 0.0058 \\
\hline
\end{tabular}
\end{table*} 

\begin{table*}
\centering
\caption{Stellar and surface correction parameters for LEGACY stars using the inverse-cubic correction method.}
\label{tab:table_b}
\begin{tabular}{lrrrrrr}
\hline
\multicolumn{1}{c}{KIC} & \multicolumn{1}{c}{Mass}  & \multicolumn{1}{c}{Initial Z}  & \multicolumn{1}{c}{Radius} &  \multicolumn{1}{c}{Age} & \multicolumn{1}{c}{\change{$\nu_{\rm corr}(\numax) / \nu_{\rm max}$}} & \multicolumn{1}{c}{$r$} \\
& \multicolumn{1}{c}{($M_{\odot}$)} & \multicolumn{1}{c}{($10^{-2}$)} & \multicolumn{1}{c}{($R_{\odot}$)} & \multicolumn{1}{c}{(Gyr)} & \multicolumn{1}{c}{($10^{-3}$)} & \\
\hline
1435467 & 1.321 $\pm$ 0.023 & 1.553 $\pm$ 0.249 & 1.683 $\pm$ 0.018 & 2.949 $\pm$ 0.200 & 2.718 $\pm$ 1.578 & 0.9974 $\pm$ 0.0047 \\
2837475 & 1.396 $\pm$ 0.040 & 1.440 $\pm$ 0.308 & 1.611 $\pm$ 0.016 & 1.744 $\pm$ 0.249 & 2.907 $\pm$ 2.981 & 0.9977 $\pm$ 0.0055 \\
3427720 & 1.098 $\pm$ 0.020 & 1.310 $\pm$ 0.203 & 1.111 $\pm$ 0.009 & 2.784 $\pm$ 0.363 & 2.891 $\pm$ 0.189 & 0.9983 $\pm$ 0.0035 \\
3456181 & 1.565 $\pm$ 0.017 & 1.948 $\pm$ 0.219 & 2.157 $\pm$ 0.019 & 1.878 $\pm$ 0.072 & 7.625 $\pm$ 0.862 & 0.9966 $\pm$ 0.0070 \\
3632418 & 1.442 $\pm$ 0.007 & 2.134 $\pm$ 0.090 & 1.908 $\pm$ 0.008 & 2.519 $\pm$ 0.052 & 6.160 $\pm$ 0.227 & 0.9915 $\pm$ 0.0039 \\
3656476 & 1.047 $\pm$ 0.014 & 2.215 $\pm$ 0.255 & 1.296 $\pm$ 0.010 & 8.644 $\pm$ 0.483 & 2.661 $\pm$ 0.350 & 0.9926 $\pm$ 0.0055 \\
3735871 & 1.109 $\pm$ 0.029 & 1.286 $\pm$ 0.240 & 1.097 $\pm$ 0.010 & 2.154 $\pm$ 0.665 & 3.142 $\pm$ 0.367 & 0.9996 $\pm$ 0.0016 \\
4914923 & 1.091 $\pm$ 0.019 & 1.874 $\pm$ 0.228 & 1.360 $\pm$ 0.012 & 6.858 $\pm$ 0.370 & 2.952 $\pm$ 0.198 & 0.9933 $\pm$ 0.0044 \\
5184732 & 1.185 $\pm$ 0.012 & 2.799 $\pm$ 0.322 & 1.325 $\pm$ 0.011 & 4.308 $\pm$ 0.223 & 3.321 $\pm$ 0.385 & 0.9920 $\pm$ 0.0088 \\
5773345 & 1.507 $\pm$ 0.023 & 2.370 $\pm$ 0.337 & 2.017 $\pm$ 0.016 & 2.194 $\pm$ 0.129 & 5.861 $\pm$ 0.622 & 0.9975 $\pm$ 0.0054 \\
5950854 & 0.972 $\pm$ 0.022 & 1.036 $\pm$ 0.160 & 1.236 $\pm$ 0.012 & 9.073 $\pm$ 0.777 & 2.335 $\pm$ 0.257 & 0.9968 $\pm$ 0.0031 \\
6106415 & 1.074 $\pm$ 0.016 & 1.297 $\pm$ 0.212 & 1.214 $\pm$ 0.009 & 5.143 $\pm$ 0.344 & 2.987 $\pm$ 0.322 & 0.9968 $\pm$ 0.0041 \\
6116048 & 1.025 $\pm$ 0.017 & 1.028 $\pm$ 0.138 & 1.219 $\pm$ 0.010 & 6.553 $\pm$ 0.404 & 3.165 $\pm$ 0.317 & 0.9935 $\pm$ 0.0043 \\
6225718 & 1.142 $\pm$ 0.021 & 1.243 $\pm$ 0.231 & 1.224 $\pm$ 0.010 & 3.044 $\pm$ 0.299 & 3.272 $\pm$ 0.465 & 0.9973 $\pm$ 0.0031 \\
6508366 & 1.583 $\pm$ 0.010 & 2.187 $\pm$ 0.181 & 2.183 $\pm$ 0.012 & 1.862 $\pm$ 0.027 & 8.074 $\pm$ 0.425 & 0.9925 $\pm$ 0.0044 \\
6603624 & 1.010 $\pm$ 0.011 & 2.139 $\pm$ 0.232 & 1.139 $\pm$ 0.009 & 8.512 $\pm$ 0.388 & 2.549 $\pm$ 0.217 & 0.9877 $\pm$ 0.0071 \\
6679371 & 1.607 $\pm$ 0.011 & 1.790 $\pm$ 0.194 & 2.216 $\pm$ 0.018 & 1.642 $\pm$ 0.045 & 4.541 $\pm$ 2.015 & 0.9953 $\pm$ 0.0058 \\
6933899 & 1.168 $\pm$ 0.011 & 2.121 $\pm$ 0.048 & 1.576 $\pm$ 0.013 & 5.933 $\pm$ 0.234 & 4.094 $\pm$ 0.248 & 0.9737 $\pm$ 0.0080 \\
7103006 & 1.471 $\pm$ 0.024 & 1.794 $\pm$ 0.195 & 1.945 $\pm$ 0.015 & 2.162 $\pm$ 0.156 & 6.331 $\pm$ 1.418 & 0.9993 $\pm$ 0.0042 \\
7106245 & 0.939 $\pm$ 0.027 & 0.478 $\pm$ 0.105 & 1.101 $\pm$ 0.014 & 6.928 $\pm$ 0.801 & 3.517 $\pm$ 1.369 & 0.9947 $\pm$ 0.0029 \\
7206837 & 1.327 $\pm$ 0.037 & 1.856 $\pm$ 0.381 & 1.565 $\pm$ 0.014 & 2.556 $\pm$ 0.362 & 2.384 $\pm$ 1.236 & 0.9997 $\pm$ 0.0059 \\
7296438 & 1.103 $\pm$ 0.022 & 2.142 $\pm$ 0.308 & 1.369 $\pm$ 0.012 & 6.852 $\pm$ 0.545 & 2.912 $\pm$ 0.415 & 0.9974 $\pm$ 0.0044 \\
7510397 & 1.434 $\pm$ 0.014 & 2.149 $\pm$ 0.257 & 1.885 $\pm$ 0.017 & 2.535 $\pm$ 0.058 & 6.537 $\pm$ 0.655 & 1.0015 $\pm$ 0.0077 \\
7680114 & 1.075 $\pm$ 0.019 & 1.742 $\pm$ 0.215 & 1.392 $\pm$ 0.012 & 7.446 $\pm$ 0.500 & 3.754 $\pm$ 0.340 & 0.9959 $\pm$ 0.0049 \\
7771282 & 1.232 $\pm$ 0.059 & 1.416 $\pm$ 0.287 & 1.622 $\pm$ 0.023 & 3.987 $\pm$ 0.765 & 2.072 $\pm$ 3.701 & 0.9993 $\pm$ 0.0038 \\
7871531 & 0.855 $\pm$ 0.013 & 0.930 $\pm$ 0.118 & 0.876 $\pm$ 0.005 & 9.285 $\pm$ 0.910 & 2.288 $\pm$ 0.149 & 0.9971 $\pm$ 0.0034 \\
7940546 & 1.478 $\pm$ 0.008 & 2.102 $\pm$ 0.088 & 1.978 $\pm$ 0.013 & 2.313 $\pm$ 0.045 & 5.964 $\pm$ 0.104 & 1.0014 $\pm$ 0.0068 \\
7970740 & 0.840 $\pm$ 0.000 & 1.085 $\pm$ 0.046 & 0.812 $\pm$ 0.001 & 7.331 $\pm$ 0.167 & 0.815 $\pm$ 0.036 & 1.0257 $\pm$ 0.0025 \\
8006161 & 0.961 $\pm$ 0.010 & 2.592 $\pm$ 0.426 & 0.920 $\pm$ 0.007 & 5.305 $\pm$ 0.384 & 1.954 $\pm$ 0.267 & 0.9995 $\pm$ 0.0061 \\
8150065 & 1.170 $\pm$ 0.045 & 1.318 $\pm$ 0.355 & 1.382 $\pm$ 0.020 & 3.936 $\pm$ 0.661 & 4.335 $\pm$ 1.143 & 0.9982 $\pm$ 0.0022 \\
8179536 & 1.199 $\pm$ 0.042 & 1.333 $\pm$ 0.272 & 1.335 $\pm$ 0.014 & 2.955 $\pm$ 0.687 & 2.119 $\pm$ 1.714 & 0.9989 $\pm$ 0.0014 \\
8228742 & 1.441 $\pm$ 0.005 & 2.531 $\pm$ 0.221 & 1.889 $\pm$ 0.007 & 2.604 $\pm$ 0.064 & 6.318 $\pm$ 0.414 & 0.9978 $\pm$ 0.0055 \\
8379927 & 1.120 $\pm$ 0.021 & 1.215 $\pm$ 0.286 & 1.116 $\pm$ 0.010 & 2.057 $\pm$ 0.295 & 2.905 $\pm$ 0.446 & 0.9988 $\pm$ 0.0033 \\
8394589 & 1.030 $\pm$ 0.028 & 0.818 $\pm$ 0.152 & 1.156 $\pm$ 0.013 & 4.997 $\pm$ 0.637 & 2.748 $\pm$ 0.281 & 0.9977 $\pm$ 0.0020 \\
8424992 & 0.922 $\pm$ 0.021 & 1.152 $\pm$ 0.203 & 1.044 $\pm$ 0.010 & 9.852 $\pm$ 1.032 & 3.041 $\pm$ 0.751 & 0.9971 $\pm$ 0.0026 \\
8694723 & 1.129 $\pm$ 0.016 & 0.935 $\pm$ 0.115 & 1.523 $\pm$ 0.016 & 4.980 $\pm$ 0.174 & 4.477 $\pm$ 0.610 & 0.9852 $\pm$ 0.0084 \\
8760414 & 0.840 $\pm$ 0.001 & 0.433 $\pm$ 0.017 & 1.034 $\pm$ 0.001 & 11.64 $\pm$ 0.112 & 2.525 $\pm$ 0.043 & 0.9972 $\pm$ 0.0006 \\
8938364 & 0.971 $\pm$ 0.014 & 1.307 $\pm$ 0.206 & 1.340 $\pm$ 0.011 & 10.741 $\pm$ 0.584 & 3.285 $\pm$ 0.310 & 0.9951 $\pm$ 0.0051 \\
9025370 & 0.992 $\pm$ 0.022 & 1.145 $\pm$ 0.368 & 1.005 $\pm$ 0.011 & 4.630 $\pm$ 0.517 & 2.550 $\pm$ 0.630 & 0.9994 $\pm$ 0.0035 \\
9098294 & 0.977 $\pm$ 0.018 & 1.145 $\pm$ 0.151 & 1.141 $\pm$ 0.010 & 8.135 $\pm$ 0.688 & 2.692 $\pm$ 0.199 & 0.9953 $\pm$ 0.0047 \\
9139151 & 1.133 $\pm$ 0.024 & 1.561 $\pm$ 0.279 & 1.141 $\pm$ 0.010 & 2.454 $\pm$ 0.495 & 3.165 $\pm$ 0.383 & 0.9993 $\pm$ 0.0021 \\
9139163 & 1.347 $\pm$ 0.033 & 1.805 $\pm$ 0.419 & 1.534 $\pm$ 0.017 & 2.070 $\pm$ 0.218 & 4.558 $\pm$ 1.047 & 0.9966 $\pm$ 0.0079 \\
9206432 & 1.385 $\pm$ 0.040 & 1.943 $\pm$ 0.407 & 1.503 $\pm$ 0.015 & 1.486 $\pm$ 0.327 & 0.823 $\pm$ 3.201 & 0.9989 $\pm$ 0.0017 \\
9353712 & 1.585 $\pm$ 0.011 & 2.191 $\pm$ 0.191 & 2.180 $\pm$ 0.014 & 1.843 $\pm$ 0.039 & 9.714 $\pm$ 0.495 & 0.9955 $\pm$ 0.0068 \\
9410862 & 0.969 $\pm$ 0.028 & 0.760 $\pm$ 0.144 & 1.150 $\pm$ 0.012 & 7.306 $\pm$ 0.916 & 1.294 $\pm$ 0.327 & 0.9979 $\pm$ 0.0021 \\
9414417 & 1.454 $\pm$ 0.015 & 1.909 $\pm$ 0.216 & 1.929 $\pm$ 0.016 & 2.331 $\pm$ 0.116 & 6.527 $\pm$ 1.053 & 0.9969 $\pm$ 0.0064 \\
9812850 & 1.391 $\pm$ 0.027 & 1.479 $\pm$ 0.188 & 1.800 $\pm$ 0.020 & 2.410 $\pm$ 0.135 & 6.912 $\pm$ 0.375 & 0.9979 $\pm$ 0.0066 \\
9955598 & 0.903 $\pm$ 0.013 & 1.492 $\pm$ 0.300 & 0.886 $\pm$ 0.007 & 6.897 $\pm$ 0.676 & 1.554 $\pm$ 0.258 & 0.9990 $\pm$ 0.0047 \\
9965715 & 1.085 $\pm$ 0.047 & 0.740 $\pm$ 0.257 & 1.268 $\pm$ 0.022 & 4.224 $\pm$ 0.701 & 3.566 $\pm$ 0.979 & 0.9985 $\pm$ 0.0014 \\
10068307 & 1.551 $\pm$ 0.022 & 2.630 $\pm$ 0.293 & 2.106 $\pm$ 0.032 & 2.060 $\pm$ 0.080 & 7.167 $\pm$ 0.399 & 0.9912 $\pm$ 0.0152 \\
10079226 & 1.104 $\pm$ 0.044 & 1.820 $\pm$ 0.339 & 1.140 $\pm$ 0.015 & 3.748 $\pm$ 1.277 & 3.431 $\pm$ 0.892 & 0.9993 $\pm$ 0.0014 \\
10162436 & 1.498 $\pm$ 0.009 & 2.130 $\pm$ 0.072 & 2.027 $\pm$ 0.012 & 2.245 $\pm$ 0.049 & 8.430 $\pm$ 0.285 & 0.9841 $\pm$ 0.0063 \\
10454113 & 1.185 $\pm$ 0.028 & 1.519 $\pm$ 0.227 & 1.242 $\pm$ 0.012 & 2.430 $\pm$ 0.438 & 2.830 $\pm$ 0.303 & 0.9935 $\pm$ 0.0054 \\
10516096 & 1.065 $\pm$ 0.018 & 1.257 $\pm$ 0.200 & 1.395 $\pm$ 0.011 & 6.903 $\pm$ 0.413 & 3.175 $\pm$ 0.344 & 0.9954 $\pm$ 0.0036 \\
10644253 & 1.147 $\pm$ 0.022 & 1.616 $\pm$ 0.281 & 1.110 $\pm$ 0.010 & 1.482 $\pm$ 0.400 & 2.920 $\pm$ 0.338 & 0.9988 $\pm$ 0.0031 \\
10730618 & 1.374 $\pm$ 0.043 & 1.712 $\pm$ 0.582 & 1.769 $\pm$ 0.028 & 2.612 $\pm$ 0.309 & 4.951 $\pm$ 1.590 & 0.9967 $\pm$ 0.0049 \\
10963065 & 1.056 $\pm$ 0.022 & 0.996 $\pm$ 0.171 & 1.214 $\pm$ 0.011 & 5.199 $\pm$ 0.442 & 3.172 $\pm$ 0.403 & 0.9974 $\pm$ 0.0032 \\
11081729 & 1.295 $\pm$ 0.052 & 1.702 $\pm$ 0.361 & 1.419 $\pm$ 0.016 & 2.113 $\pm$ 0.661 & 2.534 $\pm$ 3.533 & 0.9995 $\pm$ 0.0018 \\
11253226 & 1.373 $\pm$ 0.037 & 1.281 $\pm$ 0.251 & 1.585 $\pm$ 0.015 & 1.808 $\pm$ 0.228 & 2.868 $\pm$ 2.226 & 0.9991 $\pm$ 0.0032 \\
11772920 & 0.852 $\pm$ 0.013 & 1.229 $\pm$ 0.382 & 0.853 $\pm$ 0.006 & 9.211 $\pm$ 0.936 & 2.015 $\pm$ 0.311 & 1.0006 $\pm$ 0.0039 \\
12009504 & 1.153 $\pm$ 0.028 & 1.302 $\pm$ 0.246 & 1.386 $\pm$ 0.013 & 4.334 $\pm$ 0.404 & 4.125 $\pm$ 0.632 & 0.9963 $\pm$ 0.0028 \\
12069127 & 1.625 $\pm$ 0.009 & 2.132 $\pm$ 0.118 & 2.291 $\pm$ 0.007 & 1.737 $\pm$ 0.047 & 8.655 $\pm$ 0.249 & 0.9971 $\pm$ 0.0031 \\
12069424 & 1.035 $\pm$ 0.009 & 1.690 $\pm$ 0.013 & 1.200 $\pm$ 0.007 & 7.333 $\pm$ 0.331 & 2.929 $\pm$ 0.161 & 0.9889 $\pm$ 0.0047 \\
12069449 & 0.990 $\pm$ 0.018 & 1.600 $\pm$ 0.150 & 1.089 $\pm$ 0.012 & 7.683 $\pm$ 0.497 & 2.588 $\pm$ 0.051 & 0.9899 $\pm$ 0.0074 \\
12258514 & 1.177 $\pm$ 0.012 & 1.800 $\pm$ 0.191 & 1.558 $\pm$ 0.011 & 5.260 $\pm$ 0.221 & 3.717 $\pm$ 0.412 & 0.9914 $\pm$ 0.0066 \\
12317678 & 1.405 $\pm$ 0.012 & 1.098 $\pm$ 0.068 & 1.823 $\pm$ 0.006 & 2.123 $\pm$ 0.065 & 4.835 $\pm$ 0.853 & 1.0004 $\pm$ 0.0034 \\
Sun & 0.993 $\pm$ 0.010 & 1.350 $\pm$ 0.001 & 0.996 $\pm$ 0.007 & 4.742 $\pm$ 0.348 & 2.407 $\pm$ 0.113 & 1.0000 $\pm$ 0.0058 \\
\hline
\end{tabular}
\end{table*} 

\label{lastpage}
\end{document}